\documentclass{article}
\usepackage{arxiv}
\usepackage{graphicx,amsmath,amsfonts,amscd,amsthm,amssymb,pstricks}
\usepackage{graphicx,color}
\usepackage{dcolumn}
\usepackage{bm}
\usepackage{longtable}
\usepackage{amsmath}
\usepackage{mathrsfs}
\usepackage{amsfonts}
\usepackage{textcomp}
\usepackage{esvect}
\usepackage{float}
\usepackage[USenglish,british]{babel}
\usepackage{environ}
\usepackage{xifthen}
\usepackage{xargs}			
\usepackage{hyperref}
\usepackage{physics}
\usepackage{multirow}

\newcommandx*{\Dif}[2][1]{\ensuremath{{\textrm{d}}^{{#1}} {#2}\,}}
\newcommandx*{\Diff}[8][1, 4, 5, 6, 7, 8]{\ensuremath{
		\ifthenelse{\isempty{#4} \AND \isempty{#5}}
			{\frac{\Dif[#1]{#2}}{{\Dif{#3}}^{{#1}}}}
			{\frac{\Dif[#1]{#2}}{
				\ifthenelse{\isempty{#4}} {\Dif{#3}} {\left(\Dif{#3}\right)^{#4}}
				\ifthenelse{\isempty{#5}} {} { \ifthenelse{\isempty{#6}} {\Dif{#5}} {\left(\Dif{#5}\right)^{#6}}}
				\ifthenelse{\isempty{#7}} {} { \ifthenelse{\isempty{#8}}{\Dif{#7}}{\left(\Dif{#7}\right)^{#8}}}
}}}}

\newenvironment{subalign}[1][]{%
	\subequations
	\ifthenelse{\isempty{#1}}{\relax}{\label{#1}}
	\align
}{%
	\endalign
	\endsubequations
}

\usepackage{comment}

\newcommand{\ie}{i.e.\ }

\newcommand{\eg}{e.g.\ }

\newcommand{\lhs}{l.h.s.\ }
\newcommand{\rhs}{r.h.s.\ }
\newcommand{\ep}{\, .}
\newcommand{\ec}{\, ,}

\allowdisplaybreaks[1]

\title{Advances on the modelling of the time evolution of dynamic aperture of hadron circular accelerators}

\author{A.~Bazzani \\
Physics and Astronomy Department, Bologna University and INFN-Bologna \\
\And
M. Giovannozzi\thanks{Corresponding author: massimo.giovannozzi@cern.ch} \\
Beams Department, CERN, 1211 Geneva 23, Switzerland \\
\And 
E.H.~Maclean \\
Beams Department, CERN, 1211 Geneva 23, Switzerland \\
University of Malta, Msida, Malta
\And
C.E. Montanari \\
Physics and Astronomy Department, Bologna University and INFN-Bologna \\
\And
F.F. Van~der~Veken \\
Beams Department, CERN, 1211 Geneva 23, Switzerland \\
University of Malta, Msida, Malta \\
\And
W. Van Goethem \\
Beams Department, CERN, 1211 Geneva 23, Switzerland \\
Departement Fysica, Universiteit Antwerpen, Antwerp, B-2020, Belgium}

\begin{document}
\maketitle
\begin{abstract}
Determining a model for the time scaling of the dynamic aperture of a circular accelerator is a topic of strong interest and intense research efforts in accelerator physics. The motivation arises in the possibility of finding a method to reliably extrapolate the results of numerical simulations well beyond what is currently possible in terms of CPU time. In earlier work, a proposal for a model based on Nekhoroshev theorem and Kolmogorov--Arnold--Moser theory was made. This model has been studied in detail and proved successful in describing the evolution of the dynamic aperture in numerical simulations, however a number of shortcomings had been identified and new models are proposed in this paper, which solve the observed issues. The new models have been benchmarked against numerical simulations for a simple system, the 4D H\'enon map, as well as a realistic, non-linear representation of the beam dynamics in the LHC at $6.5$~TeV providing in both cases excellent results.
\end{abstract}



\section{Introduction}
The advent of superconducting, high-energy hadron colliders elevated non-linear beam dynamics to the forefront of accelerator design and operation. In the domain of single-particle dynamics, the concept of dynamic aperture (DA), namely the extent of the phase space region where bounded motion occurs, has been the key observable to guide the design of several past (see, \eg~\cite{tev1,tev2,tev3,herap1,herap2,herap3,rhic}), present (see, \eg~\cite{LHCDR}), and future hadron machines (see \eg ~\cite{fut1,fut2,fut3,fut4,fut5,TDR,fut6,fut7}). 

DA involves a number of challenging aspects, notably to understand which mechanism are determining its features as well as to address the numerous computational issues. In this paper we focus on a very specific aspect, namely the possibility of modelling the scaling law of DA as a function of the number of turns. This problem has been considered since the end of the $90$s~\cite{dynap1,invlog}, as finding how to describe and efficiently predict the value of the DA would solve some fundamental problems in accelerator physics and performance optimisation of circular accelerators. A reliable model for the time evolution of the DA would allow the severe limitations in terms of CPU-time to be overcome. In fact, to perform numerical simulations required to predict the performance of a circular accelerator over a realistic time interval is beyond the reach of current computers. For the LHC case, simulations up to $10^6$ turns are at the limit of the CPU-time capabilities, but this represents only $\approx 89$~s of storage time, whereas a typical fill time is of the order of several hours. Ultimately, such a model would also open the possibility to study observables that are more directly linked with machine's performance, such as beam losses and lifetime~\cite{da_and_losses}. 

To obtain a satisfactory solution to this problem, the attempt made in earlier work addressed the possibility of building models for the DA scaling with time based on fundamental results of dynamical system theory, such as Kolmogorov--Arnold--Moser (KAM)~\cite{KAM1,KAM2,KAM3,KAM4} theory and Nekhoroshev~\cite{Nekhoroshev:1971aa,Nekhoroshev:1977aa,Bazzani:1990aa,Turchetti:1990aa} theorem. In fact, to ensure applicability across machines and for different physical conditions, we need a scaling law based on the solid ground of fundamental theorems. Although very successful, this approach revealed two issues: the possibility of obtaining nonphysical model's parameters~\cite{invlog,da_and_losses} and the presence of internal dependencies among them~\cite{LumiI,LumiII}. The first generates a contradiction with the key assumptions of the scaling law, as, according to KAM and Nekhoroshev theorems, the parameters should always be positive. The latter affects the numerical stability of the model parameters whenever their dependence on physical quantities, such as linear coupling, is investigated, practically preventing this type of investigations. 

An in-depth review has been carried out, the outcome of which is that it is indeed possible to overcome the two limitations observed. The first is solved by proposing a scaling law based on Nekhoroshev theorem, only. This is justified by the fact that the condition for the applicability of the  stability-time estimate provided by Nekhoroshev theorem is more general than the existence conditions of KAM tori. Moreover, the phenomenon of Arnold diffusion that occurs in generic Hamiltonian systems with more than two degrees of freedom is extremely slow and affects a set of initial condition having a very small measure. The second one is solved by using the parameters' dependencies obtained from a closer inspection of the form of the estimate of the stability time of Nekhoroshev theorem. All these important advances will be presented and discussed in detail in this paper. 

{ The applicability of the Nekhoroshev theorem to circular accelerators requires two assumptions, namely that the system under consideration is time independent and that it is quasi-integrable with analytic dependence on the phase space variables. The first assumption can be always satisfied by extending the phase space of time-dependent systems. The latter assumption is not satisfied closed to the DA. Nonetheless, the functional form of the stability-time estimate provided by the Nekhoroshev theorem is a very robust result. In fact, it is based on the optimal estimate of the remainders of the perturbative series rather than on their convergence properties. Therefore, it is applicable even when a large fraction of the KAM tori are destroyed and replaced by a weakly-chaotic region.}

The plan of the paper is as follows: in Section~\ref{sec: Nekhoroshev and Lambert} the estimate of the stability time derived from the Nekhoroshev theorem is briefly reviewed and the connection with the Lambert-$\mathcal{W}$ function is discussed. The models proposed for describing the time variation of the dynamic aperture are presented and reviewed in detail in Section~\ref{sec:DAmodel}, while their behaviour is analysed in Section~\ref{sec:DAcomp}. Here, applications to the results of numerical simulations for the 4D H\'enon map and a detailed model of the beam dynamics in the LHC at $6.5$~TeV are presented. { In Section~\ref{sec:predictivity} the predictive power of the proposed DA models is discussed in some detail, while} an interesting observation of the properties of the DA models is discussed in Section~\ref{sec:digression}. Finally, the conclusions are drawn in Section~\ref{sec:conc} and some detailed considerations on the Lambert function can be found in the Appendices.
\section{Nekhoroshev theorem and the Lambert-$\mathcal{W}$ function} \label{sec: Nekhoroshev and Lambert}
The Nekhoroshev theorem provides an estimate for the number of turns $N(r)$ for which the orbit of an initial condition of amplitude $r$ remains bounded~\cite{Nekhoroshev:1977aa,Bazzani:1990aa,Turchetti:1990aa}, namely
\begin{equation}\label{NekhoroshevTh}
	\frac{N(r)}{N_0} =
		\sqrt{\frac{r}{r_\ast}}\exp{\left(\frac{r_\ast}{r}\right)^{\frac{1}{\kappa}}}
		\ec
\end{equation}
where $r_\ast$ and $\kappa$ are positive quantities { each capturing some key features of the system under consideration.  In~\cite{Bazzani:1990aa} the estimate of the stability time has been obtained by firstly re-scaling all co-ordinates by appropriate factors so that they do not have a physical dimension. $r_\ast$ is a dimensionless constant whose value represents an apparent convergence radius of the asymptotic perturbative series arising in the Normal Form problem of a symplectic map near an elliptic fixed point. According to this interpretation of the physical meaning of $r_\ast$, its value should decrease when increasing the strength of the non-linearities present in the system.

In the original version of Nekhoroshev theorem~\cite{Nekhoroshev:1977aa}, $\kappa$ is a function of the number $d$ of degrees of freedom of the system under consideration. However, there is no guarantee that the expression found for $\kappa$ represents an optimal estimate. In the case of a symplectic map near an elliptic fixed point~\cite{Bazzani:1990aa,Turchetti:1990aa} the simpler expression $\kappa \approx (d+1)/2$ was given in a generic framework, but once again without guarantee of being the optimal estimate when one considers a specific case.}

Equation~\eqref{NekhoroshevTh} is valid for the region in $r$ constrained by
\begin{equation}\label{eq: NekhoroshevRegion}
	\frac{r}{r_\ast} \leq \left(\frac{2}{3\kappa}\right)^{\!\!\kappa} \ep
\end{equation}
For the sake of generalisation, we recast Eq.~\eqref{NekhoroshevTh} into
\begin{equation}\label{eq: Nekhoroshev}
	\frac{N(r)}{N_0} =
		\left(\frac{r}{r_\ast}\right)^{\!\lambda}\exp{\left(\frac{r_\ast}{r}\right)^{\frac{1}{\kappa}}}
\end{equation}
where $\lambda \geq 0$. We can recover the original formulation (Eq.~(2.16) in \cite{Bazzani:1990aa}) by making the identifications:
\begin{subalign} 
	\lambda &=\label{parameter1}
		\frac{1}{2} \ec\\
	\rho_\ast &=\label{parameter2}
		\left(\frac{\kappa}{2\text{e}}\right)^{-\kappa} r_\ast \ec  \\
	N_0 &=\label{parameter3}
		\frac{7\sqrt{6}}{48}\, r_\ast^\lambda
\end{subalign}
from which it is clear that $r_\ast, \rho_\ast, \kappa$ and $N_0$ are not independent parameters. { In particular, it is worth stressing that $N_0$ is not and independent parameter, but if a function of $r_\ast, \lambda$. For this reason, it will not be used as a fit parameter in the models presented in the following sections.} Equation~\eqref{parameter3} is consistent with the theoretical result only for the case $\lambda=\frac{1}{2}$. For other values of $\lambda$, the scaling $r_\ast^\lambda$ still holds but the factor $\frac{7 \sqrt{6}}{48}$ might be normalised differently. When fitting models where $\lambda\neq\frac{1}{2}$, it is therefore advised to either use $N_0$ as a fitting parameter, or to ignore the normalisation factor and only use the scaling, \ie 
\begin{equation}
    N_0 \;\propto\; r_\ast^\lambda \ep
\end{equation}
\subsection{Inverting the Nekhoroshev stability-time estimate}
To be able to invert Eq.~\eqref{eq: Nekhoroshev}, we need to solve equations of the form
\begin{equation}\label{eq: LambertWabsolve}
	z = w^{\zeta_1} \exp{w^{\zeta_2}} \ep
\end{equation}

First we make the substitution $\tau = \frac{{\zeta_2}}{\zeta_1}w^{\zeta_2}$:
\begin{equation}
	z = \left( \frac{\zeta_1}{{\zeta_2}} \tau \right)^{\frac{\zeta_1}{{\zeta_2}}} \exp{\frac{\zeta_1}{{\zeta_2}} \tau} \ec
\end{equation}
and next we take the root $\zeta_1/{\zeta_2}$
\begin{equation}
	\frac{{\zeta_2}}{\zeta_1} z^{\frac{{\zeta_2}}{\zeta_1}} = \tau\,\exp{\tau} \ec
\end{equation}
which we can now trivially solve:
\begin{equation}
	\tau = \mathcal{W}\left( \frac{{\zeta_2}}{\zeta_1} z^{\frac{{\zeta_2}}{\zeta_1}} \right)
	\qquad\Rightarrow\qquad
	w = \sqrt[{\zeta_2}]{ \frac{\zeta_1}{{\zeta_2}} \mathcal{W}\left( \frac{{\zeta_2}}{\zeta_1} z^{\frac{{\zeta_2}}{\zeta_1}} \right) } \, ,
\end{equation}
where $\mathcal{W}$ is the so-called Lambert-$\mathcal{W}$ function, a multi-valued function whose properties are briefly reviewed in Appendix~\ref{app:LambertW}.
Comparing \eqref{eq: LambertWabsolve} with \eqref{eq: Nekhoroshev} and identifying $\zeta_1=\lambda$ and ${\zeta_2}= -\frac{1}{\kappa}$, we finally get:
\begin{equation} \label{eq: DANekhoroshev}
		r = r_\ast \left[
			\frac{1}{\Lambda} \,\,\mathcal{W}\!\!\:\!\left(
					\Lambda
					\left( \frac{N}{N_0} \right)^{\!\Lambda}
			\right)
		\right]^{\!-\kappa} \quad\text{ where }\Lambda=-\frac{1}{\lambda\kappa} \, .
\end{equation}

The choice between the two real branches $\mathcal{W}_0$ and $\mathcal{W}_{-1}$ (see Appendix~\ref{app:LambertW}) will be determined by the requirement that $\mathcal{W}$ remains real and it will hence depend on the sign of the parameters as shown in Appendix~\ref{app:parameters}. The summary is reported in the following table, where conditions on the parameters and the validity region are reported.
\begin{table}[htb]
\centering
	\caption{Possible values for the parameters $\lambda$ and $\kappa$, and the resulting admissible region of validity.}
	\begin{tabular}{c|c|c|c}
		$\lambda$ & $\kappa$ & Branch & Validity region \\ \hline\hline & & & \\
		$0 < \lambda \leq \frac{3}{2}$ & $\kappa > 0$ & $\mathcal{W}_{-1}$ &
		$\displaystyle
			\frac{N}{N_0} \geq  \left(\frac{2}{3\kappa}\right)^{\!\!\lambda\kappa} \exp{\frac{3}{2}\kappa}$ \\ & & & \\ \hline & & & \\
		$\lambda > \frac{3}{2}$ & $\kappa > 0$ & $\mathcal{W}_{-1}$ &
		$\displaystyle
			\frac{N}{N_0} \geq \left(\frac{\text{e}}{\lambda\kappa}\right)^{\!\!\lambda\kappa}$ \\ & & & \\
	\end{tabular}
	\label{tab: ParameterValues}
\end{table}
\subsection{Application of the Lambert function and of its  series expansion}
Equation~\eqref{eq: DANekhoroshev} represents the formula linking the size of the stability region to the stability time. Therefore, the closed-form model for the scaling law of dynamic aperture is represented by
\begin{equation}
D(N) = r_\ast \left[
			\frac{1}{\Lambda} \,\,\mathcal{W}_{-1}\!\!\:\!\left(
					\Lambda
					\left( \frac{N}{N_0} \right)^{\!\Lambda}
			\right)
		\right]^{\!-\kappa} \quad \Lambda=-\frac{1}{\lambda\kappa} \ep
\label{eq: new model}
\end{equation}

While Eq.~\eqref{eq: new model} is the exact solution of the dynamic aperture scaling law according to the Nekhoroshev estimate, it might not be so useful in practice. To this aim, one might use the series expansion of $\mathcal{W}_{-1}$ as given in~\cite{Knuth}, namely
\begin{equation}
\begin{split}
\mathcal{W}_{-1}(x) & = \ln (-x) - \ln\left(- \ln\left(-x\right) \right)+ \\
& + \sum_{l=0}^\infty \sum_{m=1}^\infty c_{lm} \left [ \ln (-x) \right]^{-(l+m)} \left [ \ln \left(-\ln(-x)\right)\right ]^{m} \ec
\label{LambertSeries}
\end{split}
\end{equation}
where
\begin{equation}
c_{lm} = \frac{(-1)^l}{m!} \left [\begin{matrix}l+m \\l+1\end{matrix} \right ]
\end{equation}
and the symbol in square brackets represents a Stirling cycle number~\cite{Knuth1,Knuth2}. 

Note that the inverse-logarithm law~\cite{dynap1,invlog} (see also Section~\ref{sec:loglaw})
\begin{equation}
	r = \frac{r_\ast}{\ln^\kappa\frac{N}{N_0}}
\label{eq: original model}
\end{equation}
can be recovered by taking the limit $\lambda\to 0^+$
\begin{equation}
	\lim_{\Lambda\to -\infty} \frac{1}{\Lambda} \mathcal{W}_{-1}\left( \Lambda z^\Lambda \right) = 
		\ln z \ep
\label{limit}
\end{equation}

The expansion~\eqref{LambertSeries} can be used to prove the limit~\eqref{limit}, as it can be recast in the following form
\begin{equation}
\begin{split}
	\lim_{\Lambda\to -\infty} \frac{1}{\Lambda} \mathcal{W}_{-1}\left( \Lambda \, \exp{\Lambda \ln z} \right ) & = \\
	\lim_{\Lambda\to -\infty} \frac{1}{\Lambda} \left \{ {\phantom{\frac{1}{1}}\!\!\!\!\!} \ln \left (|\Lambda| \exp{\Lambda \ln z} \right ) \right . & +   \\
	- \ln \left(-\ln \left (|\Lambda| \exp{\Lambda \ln z} \right )\right) + \\
 + \sum_{l=0}^\infty \sum_{m=1}^\infty c_{lm}  \left [ \ln \left ( |\Lambda| \exp{\Lambda \ln z} \right ) \right ]^{-(l+m)} & \times \\
 \left . \times \left [ \ln \left (- \ln \left ( |\Lambda| \exp{\Lambda \ln z} \right ) \right ) \right ]^{m}\phantom{\frac{1}{\Lambda}}\right \} & \ep
\end{split}
\end{equation}

We note that the series expansion depends on two terms of the same form that can be transformed into
\begin{align}
\ln \left (|\Lambda| \exp{\Lambda \ln z} \right ) & = \ln |\Lambda| + \Lambda \ln z \\
\ln \left(- \ln \left (|\Lambda| \exp{\Lambda \ln z} \right) \right) & = \ln \left ( - \ln |\Lambda| +|\Lambda| \ln z \right ) \ep
\end{align}

These terms are divided by $\Lambda$, which overcompensates the logarithmic divergence of the various terms except for that of the form $\Lambda \ln z$. Hence, the result in Eq.~\eqref{limit} is easily proven.
\section{Models of dynamic aperture time evolution}\label{sec:DAmodel}
\subsection{Original model and its improvement}\label{sec:loglaw}
Based on the outcome of detailed numerical simulations for several accelerator models, in~\cite{invlog} a description of the time evolution of the dynamic aperture was proposed in the form of
\begin{equation}
\begin{split}
\textbf{Model 1} \qquad \Rightarrow \qquad D(N) & = D_\infty + \frac{b}{\ln^\kappa N} \, , 
\label{model_old}
\end{split}
\end{equation}
where $D_\infty$ represents the asymptotic value of $D(N)$ and can be justified in the framework of  KAM theory, while the $N$-dependent term is derived from Eq.~\eqref{eq: Nekhoroshev} for $\lambda=0, N_0=1$ and the fit parameters are $D_\infty, b, \kappa$. Note that $b=r_*$ from Eq.~\eqref{NekhoroshevTh}. In Ref.~\cite{invlog} it was mentioned that in some conditions the fit parameters might become negative. This implies that, strictly speaking, for those cases the scaling law cannot be justified in terms of Nekhoroshev theorem, which is not satisfactory as we would like to propose a general scaling law for dynamic aperture supported by fundamental theorems of dynamical systems theory. Moreover, in several subsequent studies a dependence between the fit parameters was observed \cite{LumiI,LumiII}. For these two reasons, an alternative form of the fit has been considered. To overcome the first limitation, in the new model the term $D_\infty$ is dropped, so that the scaling law is based only on the stability-time estimate provided by the Nekhoroshev theorem. To address the second point, the inter-dependence between the parameters~\eqref{parameter1}-\eqref{parameter3} has been taken into account.
\begin{equation}
\begin{split}
\textbf{Model 2} \qquad \Rightarrow \qquad D(N) & =\rho_\ast \left ( \frac{\kappa}{2 \text{e}} \right )^\kappa \, \frac{1}{ \ln^\kappa \frac{N}{N_0}} \ec
\label{model2.1_1}
\end{split}
\end{equation}
where the free parameters are $\rho_\ast, \kappa, N_0$. By comparing Eqs.~\eqref{model_old} and~\eqref{model2.1_1} one obtains
\begin{equation}
b=\rho_\ast \left ( \frac{\kappa}{2 \text{e}} \right )^\kappa 
\label{bvskappa}
\end{equation}
which is exactly the relation from Eq.~\eqref{parameter2} as $r_\ast=b$ and it represents a first hint to explain the observed dependence between the fit parameters in Model~1. 

Parenthetically, the model~\eqref{model2.1_1} can be written also in the following form 
\begin{equation}
\ln D(N) = \ln \rho_\ast + \kappa \left [ \ln \kappa - \ln (2\, \text{e}) -\ln \left ( \ln \frac{N}{N_0} \right ) \right ] \ec
\end{equation}
which can be more convenient for a numerical application. In this case the natural choice for the fit parameters is $\;\ln \rho_\ast, \kappa, N_0$.
\subsection{New models based on the Lambert function}
By considering the expressions for the parameters as given in Eqs.~\eqref{parameter1}-\eqref{parameter3} it is possible to recast the scaling law for the dynamic aperture~\eqref{eq: new model} in the following form
\begin{equation}
\begin{split}
& \textbf{Model 4} \qquad \Rightarrow \qquad D(N) = \rho_\ast \times \\
& \times \displaystyle{\frac{1}{\left[-2 \, \text{e} \, \lambda \,\mathcal{W}_{-1}\!\!\:\!\left(
-\frac{1}{2 \, \text{e} \, \lambda}\left( \frac{\rho_\ast}{6} \right)^{1/\kappa} \, \left( \frac{8}{7} N \right)^{-1/(\lambda \, \kappa)} \right)
		\right]^{\kappa}}} \ec \phantom{\times}
\end{split}
\label{eq: new model exact}
\end{equation}
where the free parameters are $\rho_\ast$, $\kappa$ and, possibly, $\lambda$, unless it is fixed to the value of $1/2$ according to the analytic Nekhoroshev estimate. In the rest of the paper Eq.~\eqref{eq: new model exact} will be indicated as Model~4. { The notable limit discussed in the previous section implies that Model~4 reduces to Model~2 when $\lambda \to 0^+$}

The series expansion of $\mathcal{W}_{-1}$ can be used to obtain an approximate model for the time evolution of the dynamic aperture. In fact, one can retain the lowest order terms and reduce  Model~4 to the form 
\begin{equation}
\begin{split}
& \textbf{Model 3} \qquad \Rightarrow \qquad D(N) = \rho_\ast \left ( \frac{\kappa}{2 \text{e}} \right )^\kappa \times \\
& \times  \displaystyle{\frac{1}{ \left [ \ln \frac{N}{N_0} +\lambda \kappa \left (\ln \lambda \kappa + \ln \left ( \ln \lambda \kappa + \frac{1}{\lambda \kappa} \ln \frac{N}{N_0} \right )  \right ) \right ]^\kappa}} \phantom{\times}
\end{split}
\label{eq: new model approximate}
\end{equation}
with
\begin{equation}
\ln N_0 = \ln \frac{7}{8} + \lambda \ln \frac{\rho_\ast}{6} + \lambda \, \kappa \left [\ln \kappa- \ln (2\, \text{e})  \right ] 
\end{equation}
where the relations~\eqref{parameter1}-\eqref{parameter3} have been used and, also in this case, the model parameters are $\rho_\ast$ and $\kappa$ and possibly $\lambda$. 

It is straightforward to verify that Model~3 tends to Model~2 for $\lambda \to 0$. While for Model~3 $N_0$ is a function of the other fit parameters, it is a constant for Model~2. It is worth noting that Model~3 features the same logarithmic behaviour as in Model~2, but it also includes $N$-independent terms and double-logarithmic ones, which represent an improvement in spite of the same number of parameters (if $\lambda$ is set to $1/2$). { In this respect, Model~3 can be seen as an intermediate one, between Model~2 and Model~4. Therefore, we expect a better performance than Model~2 and a simpler numerical implementation than Model~4 that requires the special function $\mathcal{W}$.} As a last remark, Model~3 can also be written in logarithmic form, which might be convenient from a numerical point of view. 
\section{Analysis of the models of dynamic aperture evolution}\label{sec:DAcomp}
In the rest of the paper, the new models of the DA evolution with time will be scrutinised by analysing their behaviour when applied to a simple dynamical system like the 4D H\'enon map and to realistic realisations of the LHC ring at top energy. The most stringent conditions are used, which means that the number of model's parameters has been reduced to the minimum, i.e. to $2$ as $\lambda$ is fixed to $1/2$ and $N_0$ is set to $1$ for Model~2 or its functional relationship on $\rho_\ast$ and $\kappa$ is applied. 
\subsection{The 4D H\'enon map}
\subsubsection{Generalities}
The 4D H\'enon map~\cite{yell} is a well-known model that combines simplicity in its form with a rich dynamical behaviour. Moreover, it represents the betatronic motion of a FODO lattice with a single sextupole in the single-kick representation. Such a system can be made more complicated by introducing a time modulation of the linear frequencies. The reason to consider the modulated version of the 4D H\'enon map in this context is twofold: on one hand the tune modulation takes into account the coupling with the longitudinal dynamics, on the other hand in~\cite{invlog} it was observed for the first time that for large values of the modulation amplitude $\epsilon$, the proposed model for the DA scaling was providing negative values of the fit parameters. 

The modulated 4D H\'enon map reads
\begin{equation}
	\mqty(x^{(n+1)} \\ p_x^{(n+1)} \\ y^{(n+1)} \\ p_y^{(n+1)}) = \vb{L} \mqty(x^{(n)} \\ p_x^{(n)} + [x^{(n)}]^2 - [y^{(n)}]^2 \\ y^{(n)} \\ p_y^{(n)} - 2x^{(n)}y^{(n)}),
\end{equation}
where $(x,p_x,y,p_y)$ are the phase-space coordinates after transformation to linear normalised co-ordinates, i.e. Courant-Snyder co-ordinates, and after rescaling by the strength of the sextupole to make the map independent from the sextupole strength~\cite{yell}. $\vb{L}$ is a matrix given by the direct product of two 2D rotations $R$, namely
\begin{equation}
	\vb{L} = \mqty(R(\omega_x^{(n)}) & 0 \\ 0 & R(\omega_y^{(n)})),
\end{equation}
where the linear frequencies $\omega_x^{(n)}$, $\omega_y^{(n)}$ are varying with the discrete time $n$ according to
\begin{align}
	\omega_x^{(n)} &= \omega_{x0}\left(1 + \epsilon \sum_{k=1}^m \epsilon_k \cos(\Omega_k n)\right),\\
	\omega_y^{(n)} &= \omega_{y0}\left(1 + \epsilon \sum_{k=1}^m \epsilon_k \cos(\Omega_k n)\right).
\end{align}

As for the values for the parameters, we have considered $\omega_{x0} = 0.168$ and $\omega_{y0} = 0.201$ and for the $\Omega_k$ frequencies and $\epsilon_k$ amplitudes the values are listed in Table~\ref{tab:henon} using the same values as in~\cite{invlog}.

\begin{table}[!htp]
	\centering
	\caption{Parameters of the modulated H\'enon map.}
	\begin{tabular}{ccc}
		\hline
		\hline
		$k$ & $\Omega_k$ & $10^4 \epsilon_k$ \\
		\hline
		1 & $2\pi / 868.12$ & $1.000$ \\
		2 & $2\Omega_1$ & $0.218$ \\
		3 & $3\Omega_1$ & $0.708$ \\
		4 & $6\Omega_1$ & $0.254$ \\
		5 & $7\Omega_1$ & $0.100$ \\
		6 & $10\Omega_1$ & $0.078$ \\
		7 & $12\Omega_1$ & $0.218$ \\
		\hline
		\hline
		\label{tab:henon}
		\end{tabular}
\end{table}

\begin{figure}[!htbp]
\centering
	\includegraphics[trim = 0mm 0mm 3mm 0mm,width=.32\linewidth,clip=]{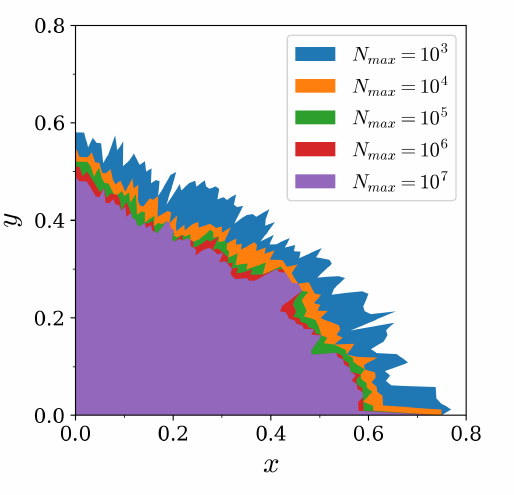}
	\includegraphics[trim = 0mm 0mm 3mm 0mm,width=.32\linewidth,clip=]{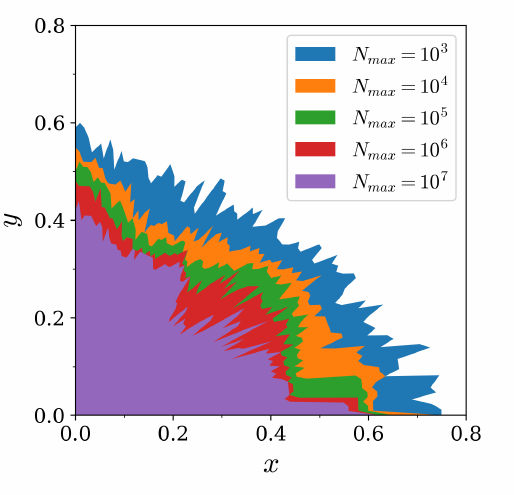}
	\includegraphics[trim = 0mm 0mm 3mm 0mm,width=.32\linewidth,clip=]{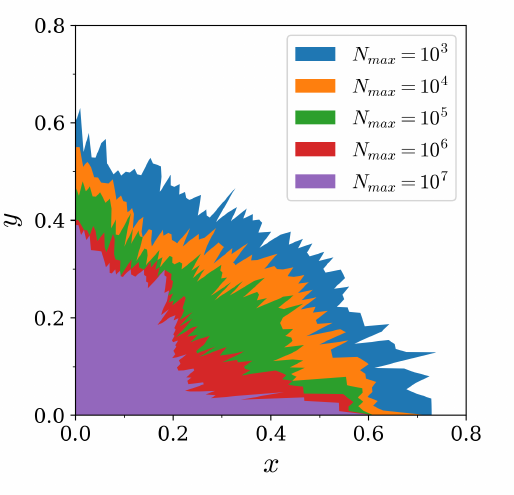}
\caption{Stability region for the modulated H\'enon map for three values of $\epsilon$, namely $0$ (left), $32$ (middle), $64$ (right). The different colours refer to different stability times $N_{\rm max}$. Note the increasing asymmetry between the horizontal and vertical planes for increasing values of $\epsilon$ and $N_{\rm max}$.}
\label{fig:henon_stab}
\end{figure}

A plot of the map stability is reported in Fig.~\ref{fig:henon_stab} for three values of $\epsilon$.

The tracking has been performed up to $10^7$ turns for several values of the modulation amplitude $\epsilon$ in the interval $0 \leq \epsilon \leq 64$. A polar grid of initial conditions have been built, with the angular interval $[0,\pi/2]$ divided into $100$ parts and a step in amplitude of $10^{-2}$. The DA of the map is computed by using the amplitude of the last stable initial condition for each angle and by taking the angular average, as discussed in~\cite{dacomp}.
\subsubsection{Results of numerical investigations}
\begin{figure}[!htp]
\centering
	\includegraphics[trim = 0mm 0mm 2mm 1.5mm,	width=.7\linewidth,clip=]{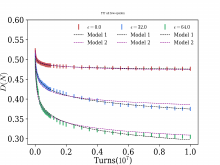}
\caption{Evolution of the dynamic aperture as a function of turn number (markers) for the three values of the $\epsilon$ parameter used in Fig.~\ref{fig:henon_stab}. The error associated with the numerical estimate of the DA are also shown, together with the results of the DA Model~1 and~2 (lines). The other two models are not shown here as they provide results very similar to those of Model~2.}
\label{fig:henon_DA}
\end{figure}

Detailed numerical simulations have been performed, aimed at computing the DA as a function of $\epsilon$ using the approach reported in~\cite{dacomp} for the the computation of the DA and the associated numerical error. A summary plot is shown in Fig.~\ref{fig:henon_DA}, where the DA as a function of turn number is reported, including the errors associated with the numerical computation, as well as the results of Model~1 and~2, only, as Model~3 and~4 provide results that are essentially indistinguishable from those of Model~2.

\begin{figure}[!htbp]
\centering
	\begin{tabular}{@{}c@{}c@{}c@{}}
	\includegraphics[trim = 13mm 5mm 11mm 25mm,width=.33\linewidth,clip=]{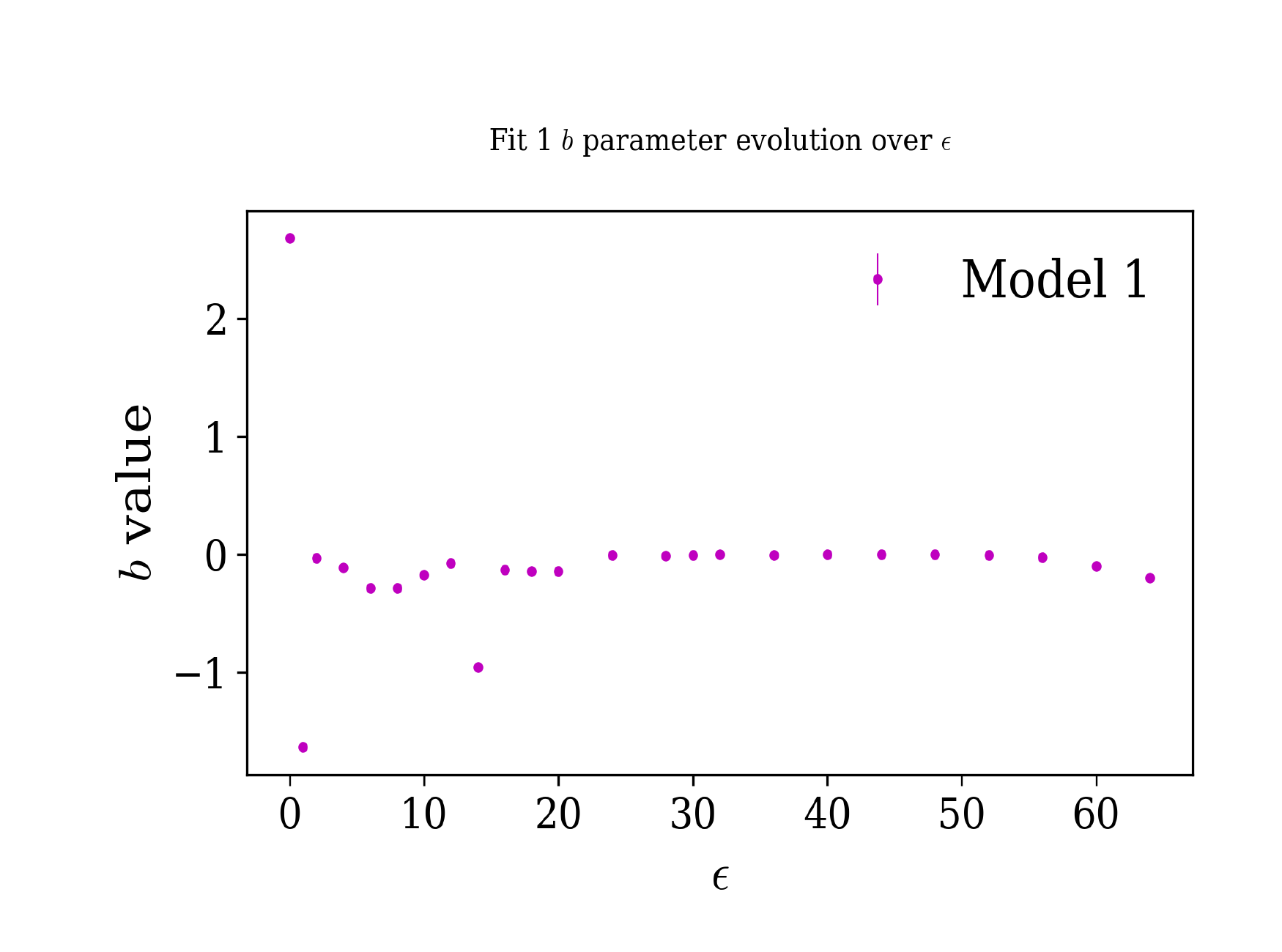} &
	\includegraphics[trim = 11mm 5mm 11mm 25mm,width=.33\linewidth,clip=]{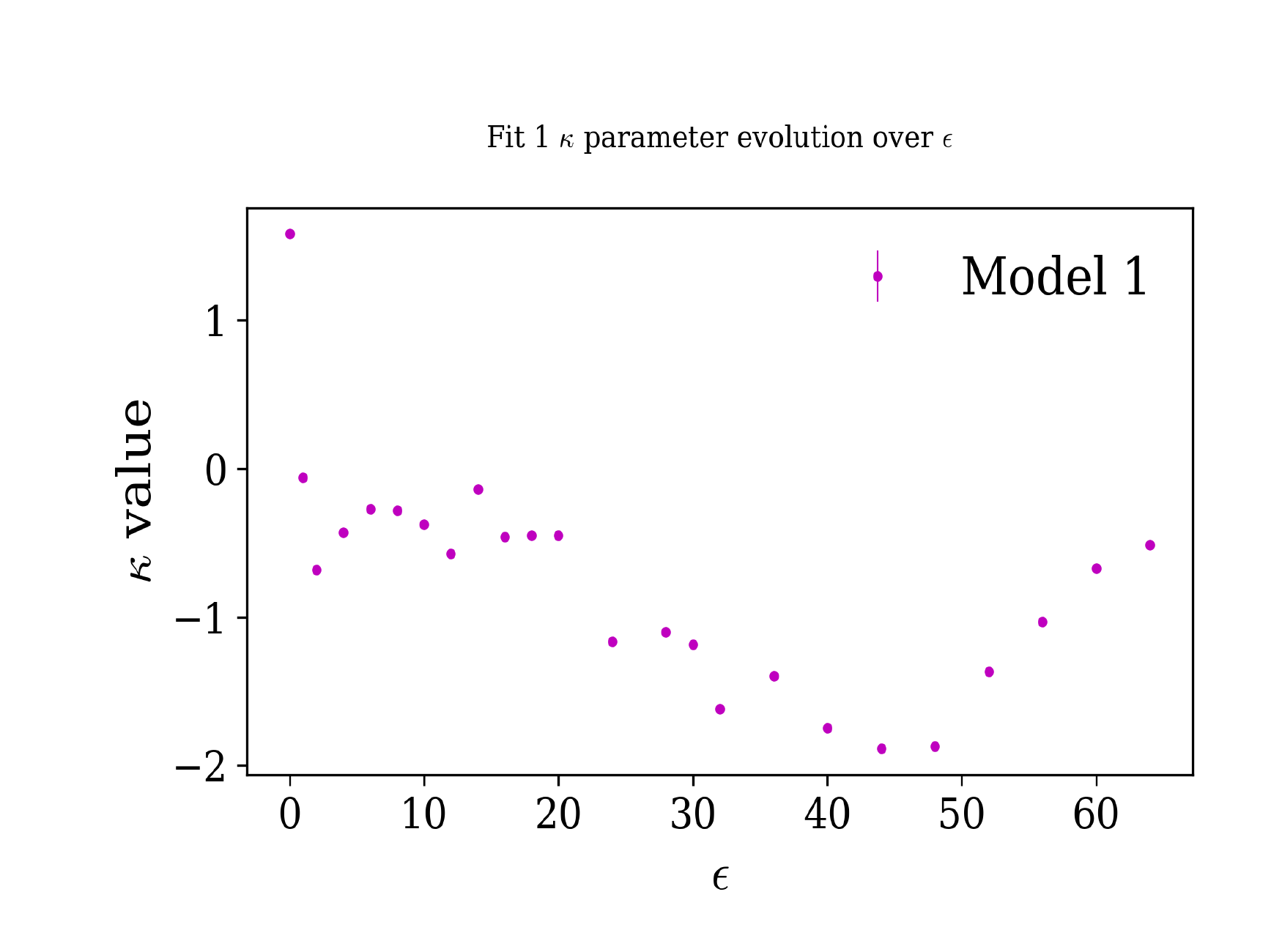} &
	\includegraphics[trim = 11mm 5mm 11mm 25mm,width=.33\linewidth,clip=]{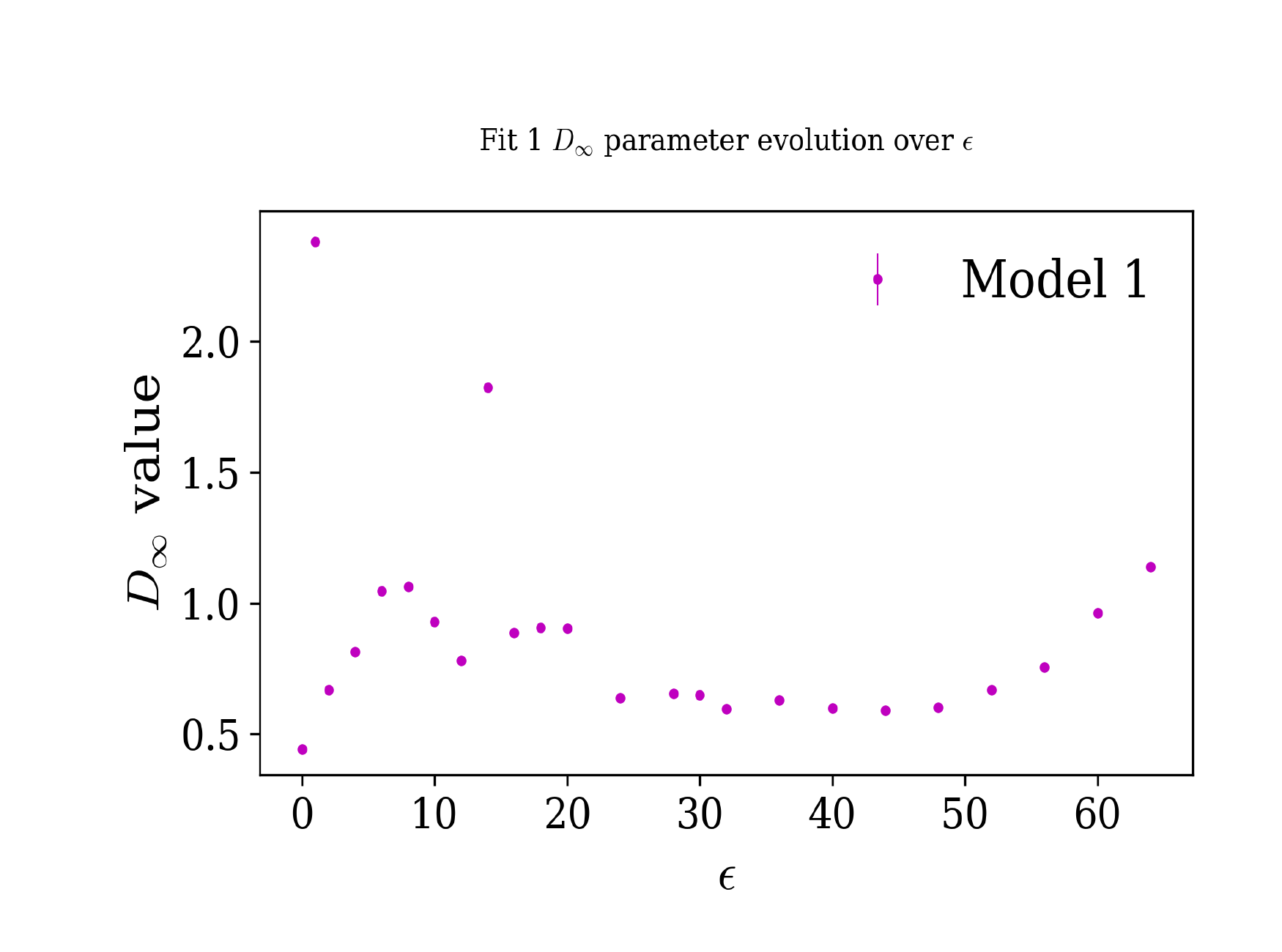} \\
	\includegraphics[trim = 13mm 5mm 11mm 25mm,width=.33\linewidth,clip=]{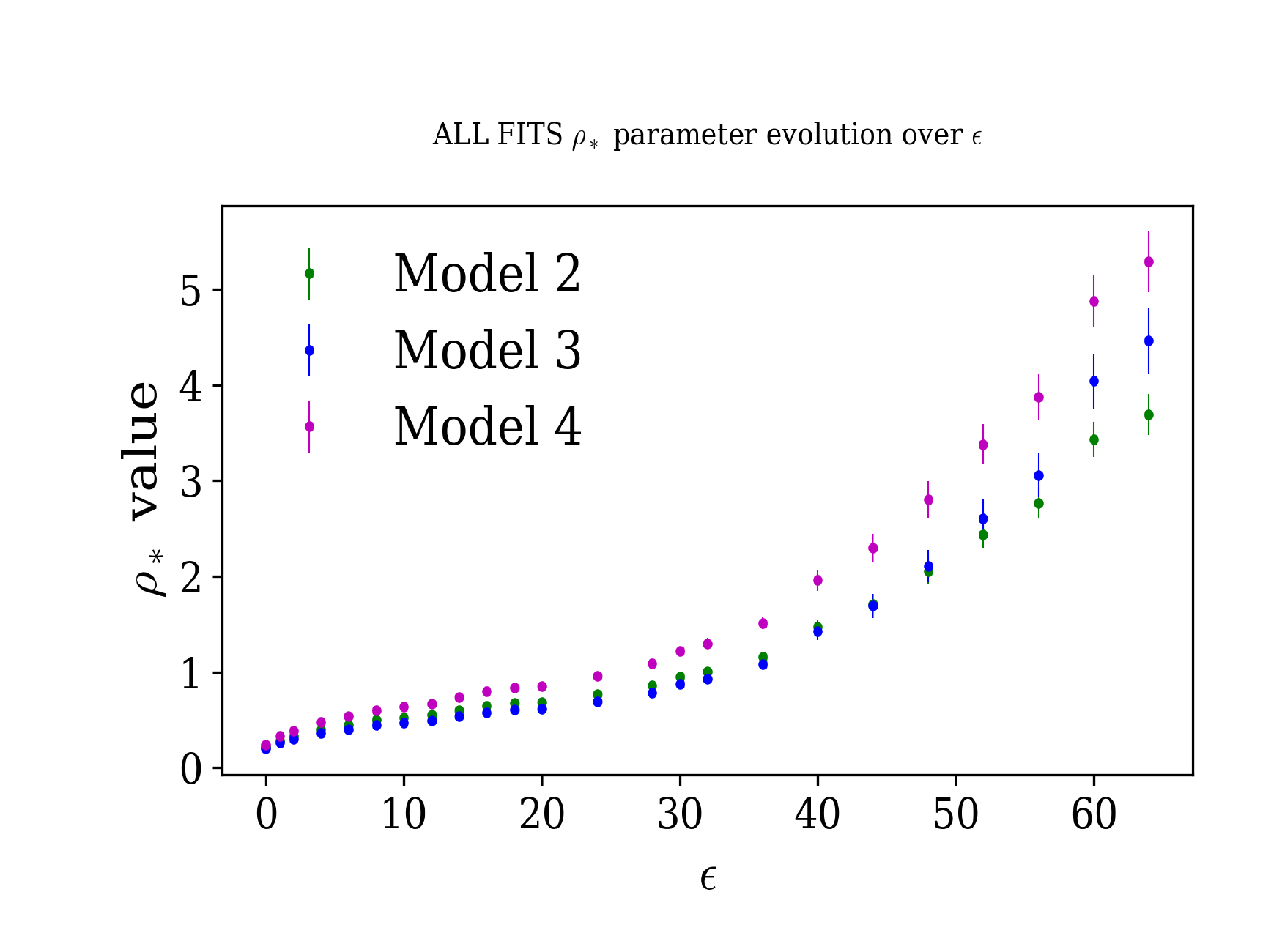} &
	\includegraphics[trim = 11mm 5mm 11mm 25mm,width=.33\linewidth,clip=]{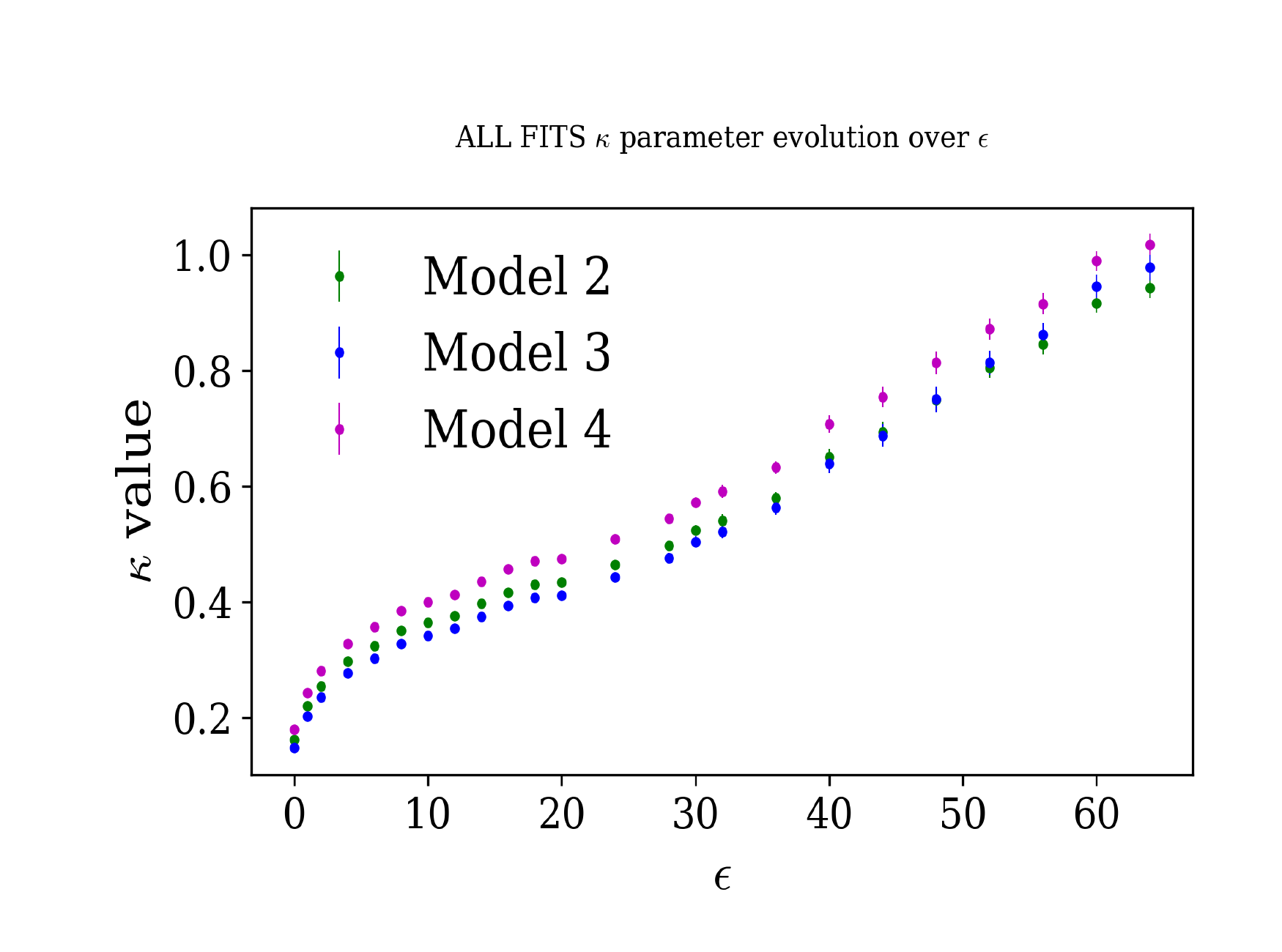} &
	\includegraphics[trim = 11mm 5mm 11mm 25mm,width=.33\linewidth,clip=]{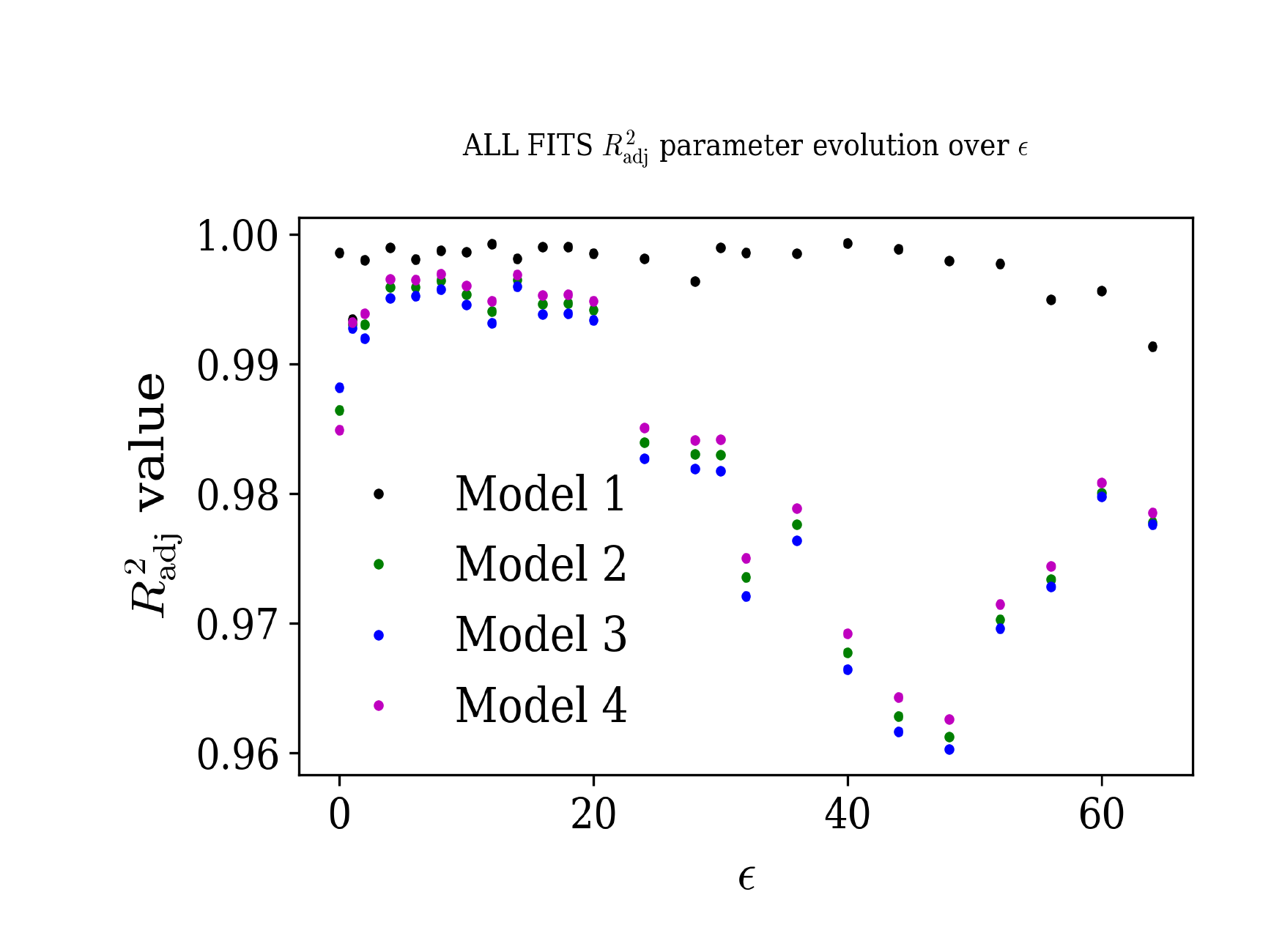} \\
	\end{tabular}
\caption{Dependence of the model parameters on $\epsilon$ for Model~1 (upper) and the other models (lower) in which $\lambda$ and $N_0$ are not free parameters. The figure of merit of the model quality, $R^2_{\rm adj}$, is also reported (lower right).}
\label{fig:henon_models}
\end{figure}

All models provide a good agreement with the numerical data. The essential differences between them can be better appreciated by inspecting their dependence on $\epsilon$, which is shown in Fig.~\ref{fig:henon_models}. In the upper row, the three parameters for Model~1 are plotted, while in the second row the two parameters of Model~2,~3, and~4 are shown, together with the $R^2_{\rm adj}$, the adjusted coefficient of determination, for all four models. For Model~1, it is clearly visible that $\kappa$ and $b$ are varying in sign, with rather larger changes of their absolute values. Such large variations are also visible in the behaviour of $D_\infty$, which also features some outliers. On the other hand, the remaining three models feature a rather smooth dependence on $\epsilon$ and a strong similarity between them. It is worth noting that Model~2 and~3 feature parameters values that resemble more to each other than those of Model~4.

These results indicate that negative parameters are only a feature of Model~1. All this indicates that the observations made in~\cite{invlog} about the presence of negative values of the model parameters are an artefact of the model form, in which a constant term $D_\infty$ is added to that derived from the Nekhoroshev estimate of the stability time. In general, this form provides positive fit parameters, but in some cases negative values might appear, which stem from the compensation between the two terms of the model. Of course, this improves the fit quality as shown in the behaviour of the $R^2_{\rm adj}$, which is in general higher for Model~1 as compared to the other models. The special care needed to analyse a model made of the sum of different terms, with respect to a model made of product of various terms, will be discussed again in Section~\ref{sec:digression}. Moreover, it is important to stress that the new models depend on two free parameters, only. Hence, they really seem to outperform Model~1. 
\subsection{Realistic LHC models}
\subsubsection{LHC dynamic aperture experiment at top energy}
In this paper, we do not aim at discussing the agreement between experimental measurements and numerical simulations, rather we use realistic models, based on experimental configurations used in the LHC, to study the performance and behaviour of the proposed models of DA variation with the number of turns. 

\begin{figure}[htb]
\begin{center}
\begin{tabular}{@{}c@{}}
{\includegraphics[trim = 4mm 0mm 5mm 2mm, width=0.6\linewidth,clip=]{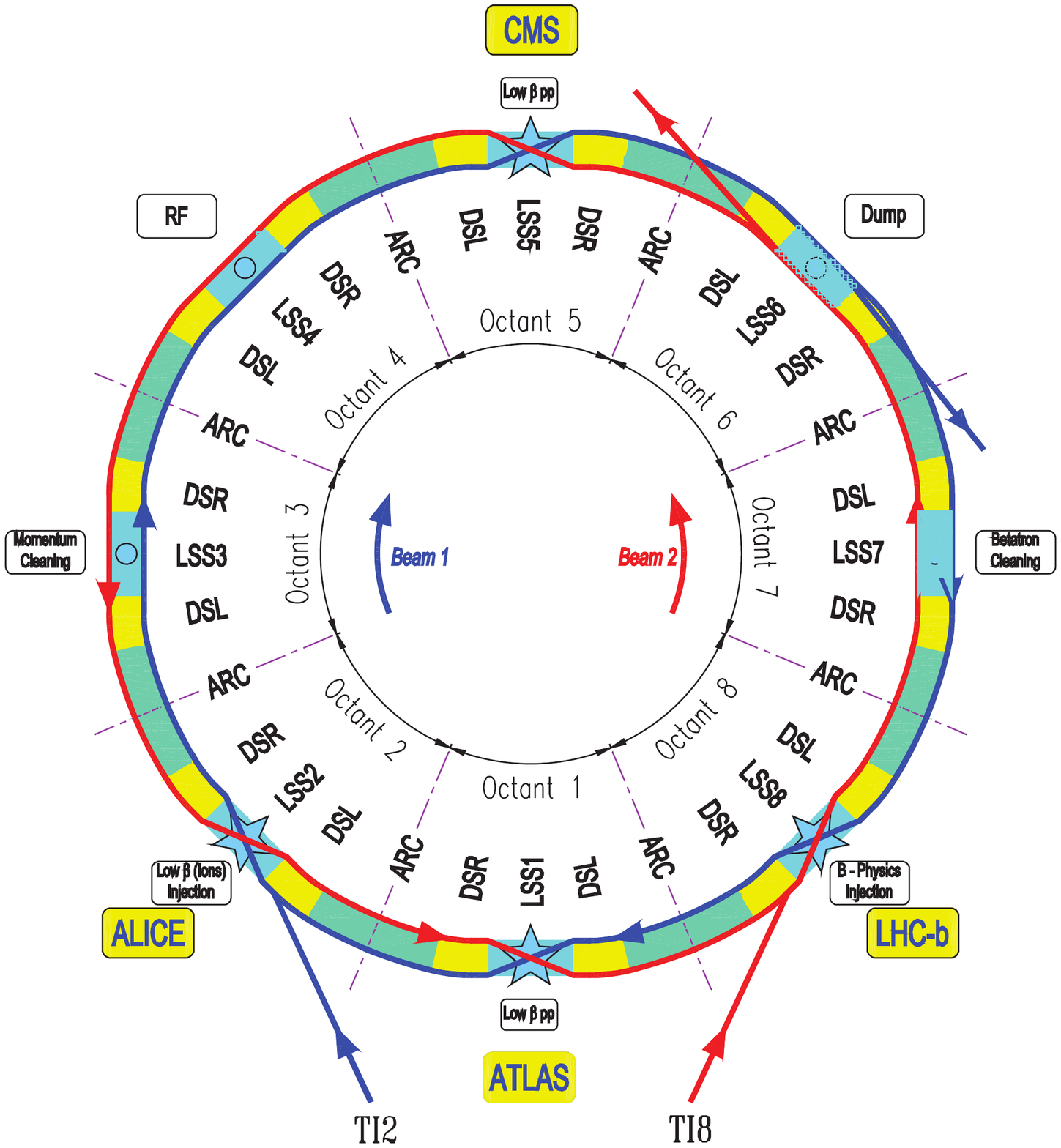}} \\
{\includegraphics[trim = 0mm 0mm 0mm 45mm, width=0.6\linewidth,clip=]{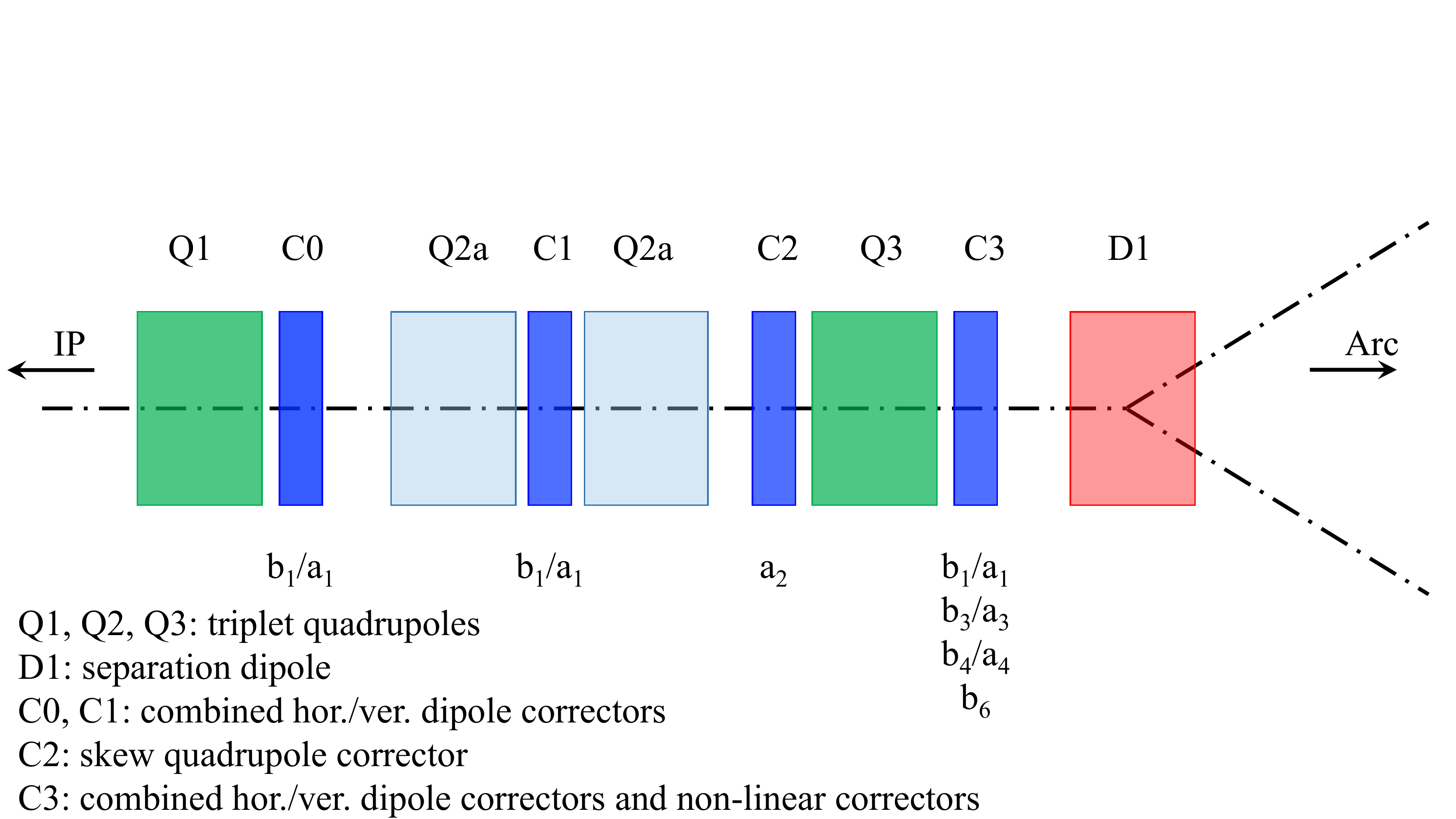}}
\end{tabular}
\caption{Upper: Layout of the LHC (from Ref.~\cite{LHCDR}). The ring eight-fold symmetry is visible, together with the arcs and the long straight sections. Bottom: Sketch of the layout of the inner triplets and the non-linear correctors used in the experimental tests reported in this paper. The field imperfections of LHC magnets are represented as $B_y + i \, B_x = B_{\rm ref} \sum_{n=1}^{M} \left ( b_n  + i \, a_n\right ) 
\left ( \frac{x + i \, y}{R_{\rm r}} \right )^{n-1}$ where $R_{\rm r}=17$~mm.}
\label{LHClayout}
\end{center}
\end{figure}

DA measurements at the LHC (see Fig.~\ref{LHClayout}, upper, for a layout of the LHC ring) have been carried out at injection energy~\cite{DABeam2,DABeam1_1,DABeam1_2} using different approaches, i.e. the standard kick method~\cite{DABeam2} or the novel approach~\cite{DABeam1_1,DABeam1_2}. 

Recently, DA measurements have been successfully performed also at at $6.5$~TeV in the LHC~\cite{MD_note}. The goals of these measurements were many-fold: the use of squeezed optics allows investigations of the impact on beam dynamics of the non-linear field errors stemming from the quadrupoles in the high-luminosity insertions. Thus, one could examine and quantify the influence on beam loss and lifetime from changes in the strength of the normal dodecapole correctors (see Fig.~\ref{LHClayout}, bottom, for a sketch of the high-luminosity insertions, whose magnets were used during the experiment) in the ATLAS and CMS interaction regions (IR) 1 and 5, respectively. This aspect is particularly relevant in view of the future High Luminosity LHC project~\cite{TDR}, for which the operational strategy to set the non-linear correctors in the high-luminosity IRs is being actively studied. 

The detail regarding the experimental session and the LHC set-up can be found in~\cite{MD_note}. Here, it is important to mention that large dodecapole sources were introduced by powering the IR-$b_{6}$ correctors left and right of the interaction point (IP) 1 and 5, uniformly to their maximum current. Finally, the IR non-linear corrections for normal and skew sextupole and normal and skew octupole errors, which had been commissioned at the 2017 start-up, were collectively removed. 

The ring model used for the numerical simulations of the DA is the most accurate description of the LHC lattice, including the measured field errors (see~\cite{DAasbuilt} for more detail) together with the operational configuration of the various correction circuits. The numerical protocol used envisages the generation of sixty realisations of the magnetic errors to take into account their measurement uncertainties. Moreover, a polar grid of initial conditions in $x-y$ space is defined and their evolution is computed for up to $10^6$ turns. The polar grid of initial conditions is obtained by dividing the first quadrant of the $x-y$ space in $59$ angles and along each direction $30$ initial conditions are uniformly distributed over intervals of $2\sigma$. The DA has been computed using the approaches described in~\cite{dacomp}. 

The evolution of the initial conditions through the LHC magnetic lattice is computed using the SixTrack code~\cite{sixtrack}, which implements a second-order symplectic integration method. All configurations used in the DA experiment at $6.5$~TeV have been simulated through { 6D} numerical simulations and they are used in the following for assessing the performance of the models  describing the time variation of the DA. Note that in the rest of the paper the configuration in which all IR correctors are powered will be indicated as Configuration~A, while that with the dodecapolar corrector only as Configuration~B.
\subsubsection{Results of numerical investigations}
\begin{figure}[!htbp]
\centering
\begin{tabular}{@{}c@{}@{}c@{}@{}c@{}c@{}}
{\includegraphics[trim = 10mm 0mm 20mm 24mm, width=0.25\linewidth,clip=]{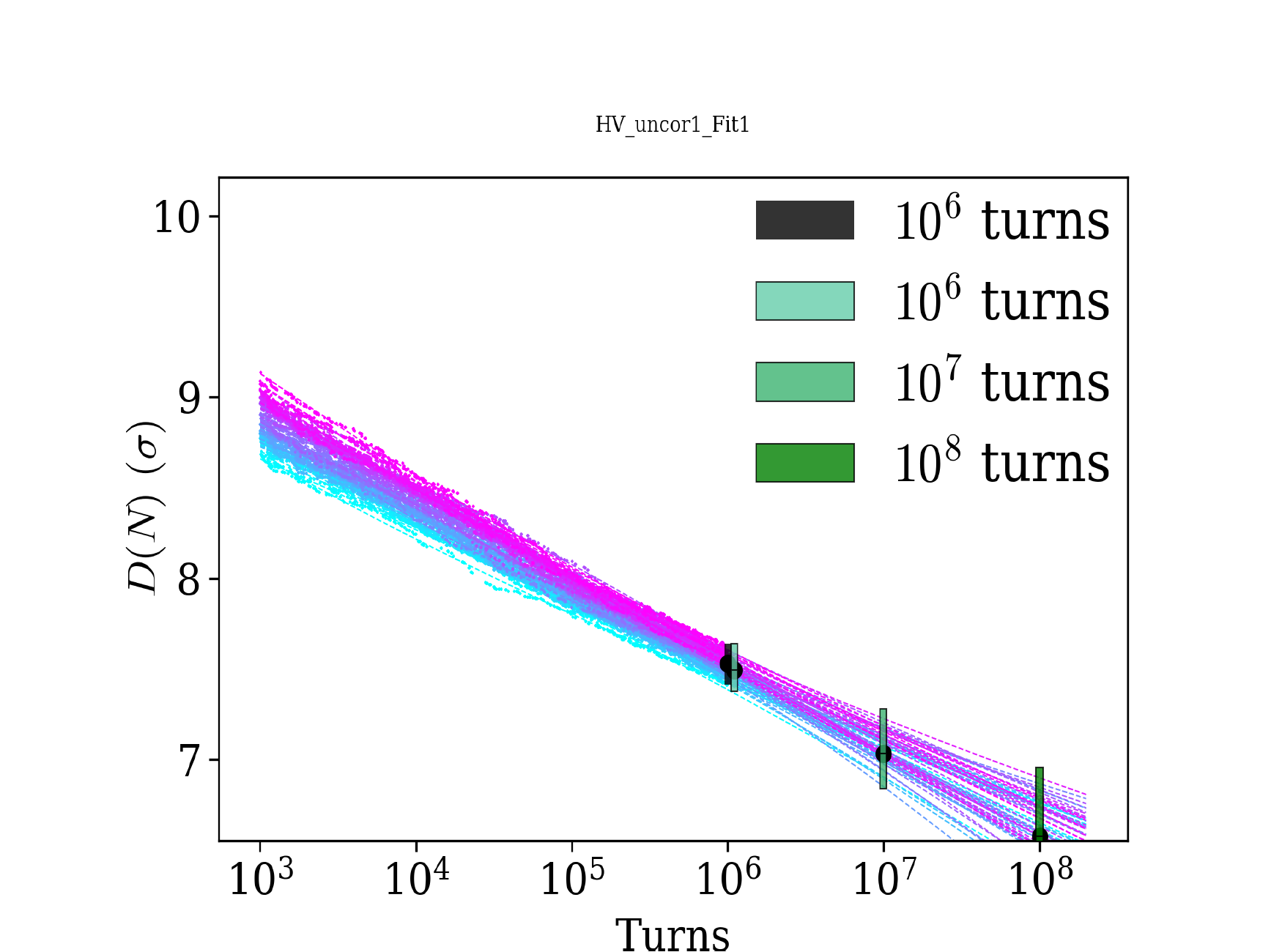}} &
{\includegraphics[trim = 10mm 0mm 20mm 24mm, width=0.25\linewidth,clip=]{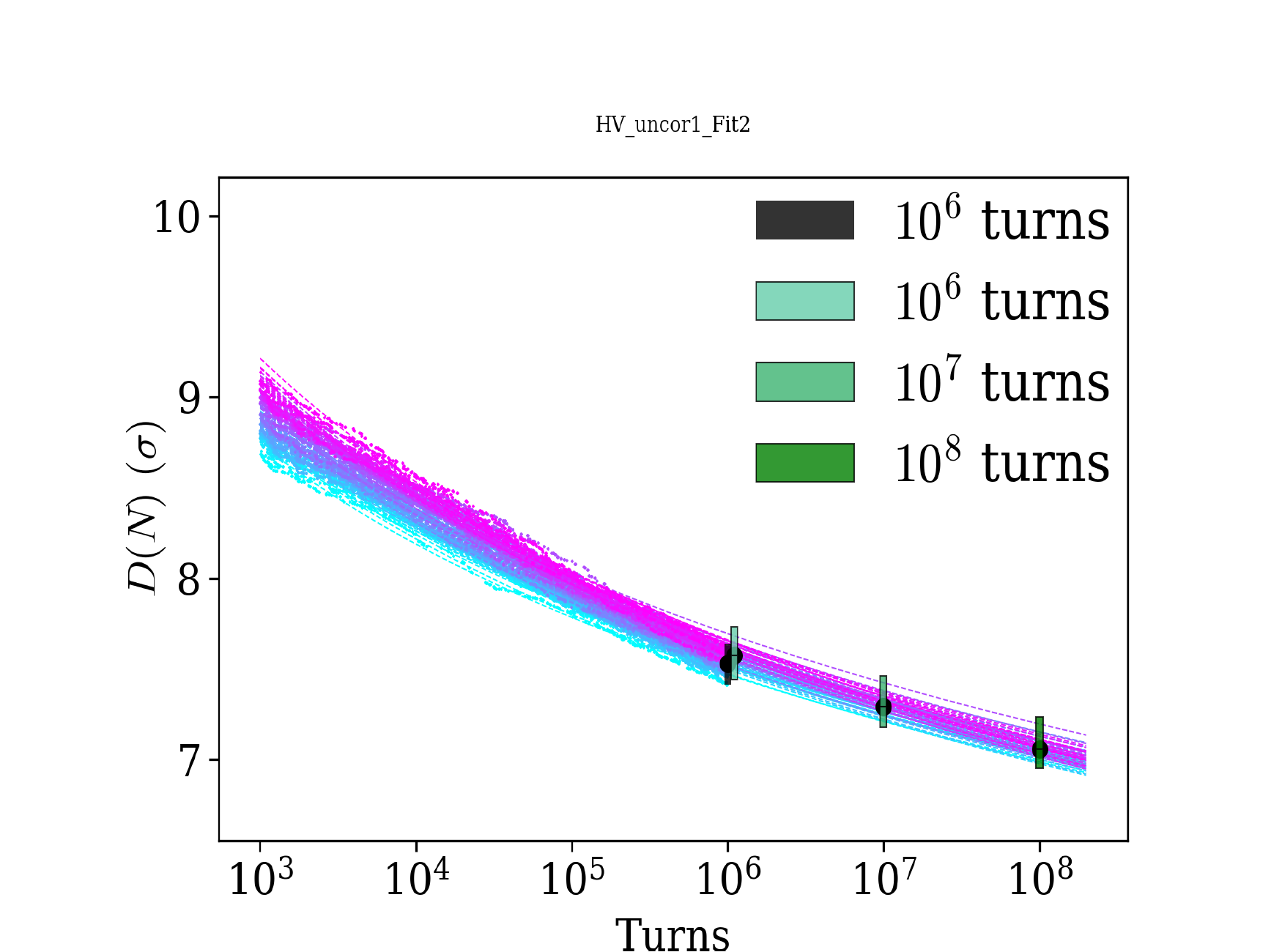}} &
{\includegraphics[trim = 10mm 0mm 20mm 24mm, width=0.25\linewidth,clip=]{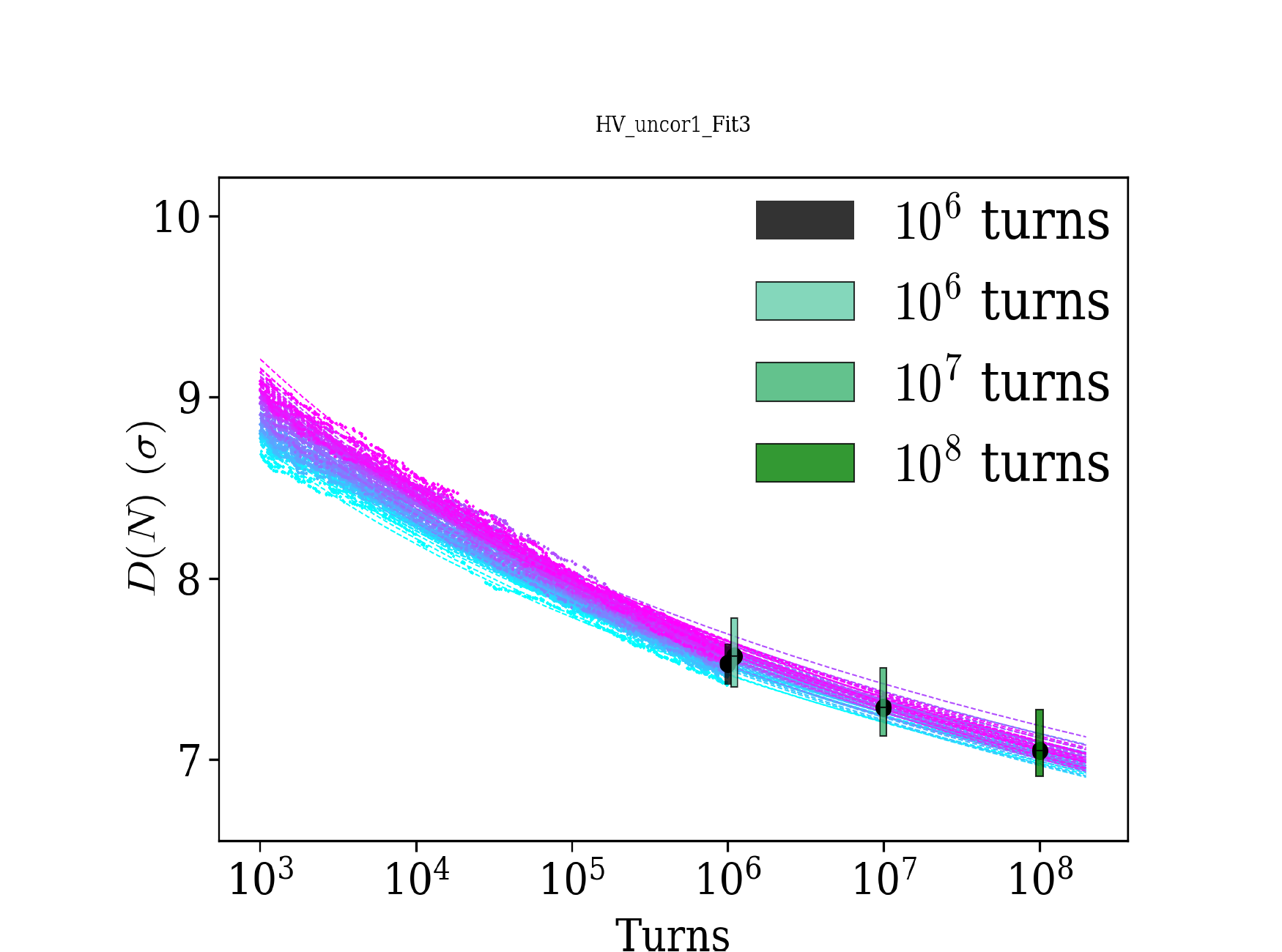}} &
{\includegraphics[trim = 10mm 0mm 20mm 24mm, width=0.25\linewidth,clip=]{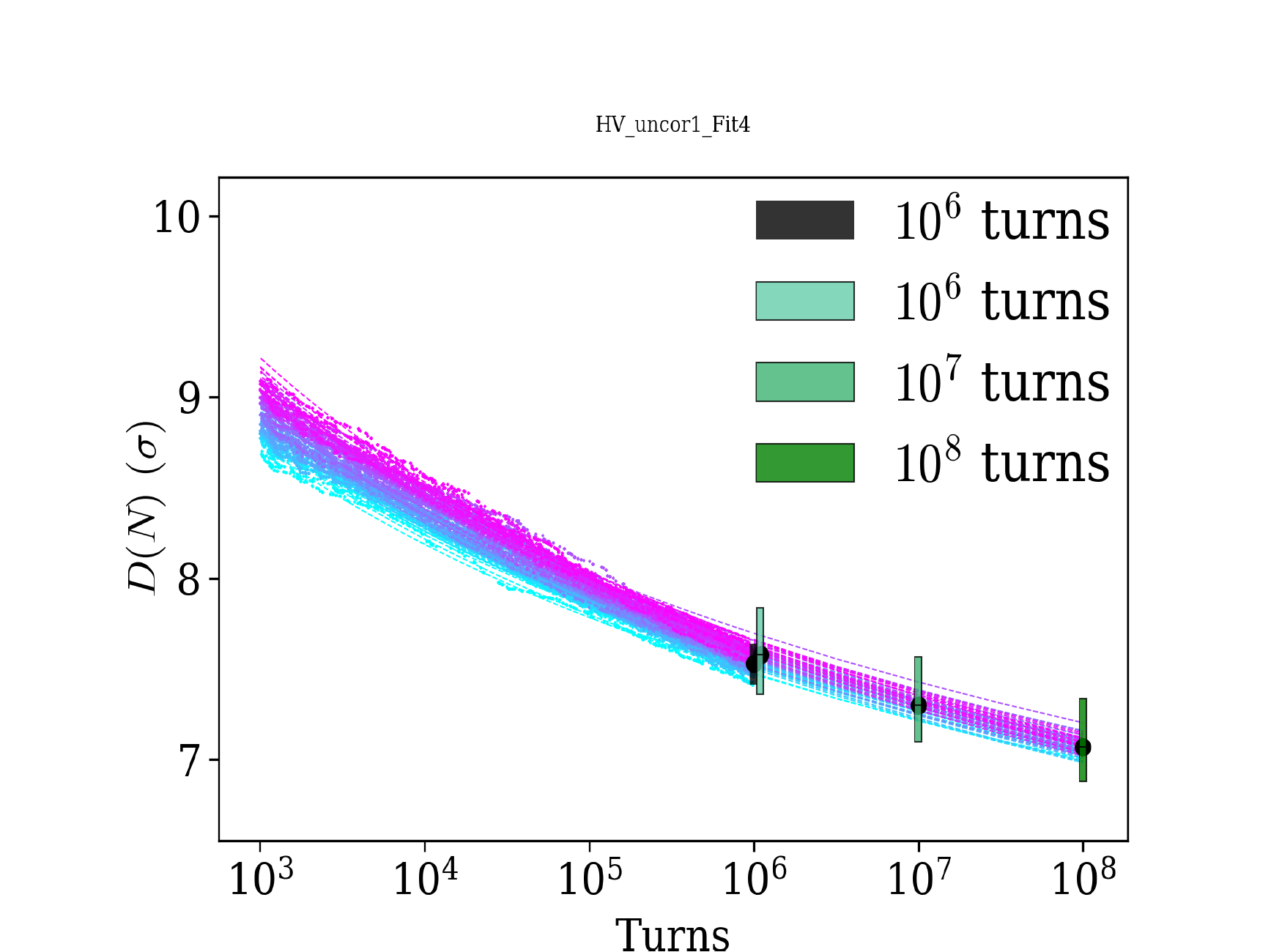}} \\
\end{tabular}
\caption{DA vs time for Configuration~B for Beam~1 for Model~1,~2,~3, and~4 (left to right, respectively). The various curves refer to the sixty realisations used in the numerical simulations, which are differentiated by their colour and stop at $10^6$ turns. Extrapolated values up to $10^8$ turns are also given, based on the models that reproduce the numerical data. The black bar indicates the DA from numerical simulations and the related uncertainty, whereas the green bars indicate the DA from the models and the related uncertainty (the corresponding numerical values are listed in Table~\ref{tab:LHCmodels}.}
\label{fig:LHCDAvstime}
\end{figure}

The comparison of the performance of the various DA models is largely independent on the LHC configuration used. Therefore, a selection has been applied and in the following the outcome of the numerical simulations for Configuration~B for Beam~1 (i.e. the clockwise beam, whereas Beam~2 is the counter-clockwise beam) will be presented and discussed in detail. Figure~\ref{fig:LHCDAvstime} shows both the DA data as computed using SixTrack for the sixty realisations (using different colours) of the LHC magnetic field errors up to $10^6$ turns. These data have been modelled using both the old model and the new ones and the corresponding curves are also shown (using the same colour palette as for the numerical data). Moreover, extrapolation of the considered models up to $10^8$ turns have been computed and the results reported, to provide quantitative information about the predictive power of the various models. 

The bars reported in the four plots are centred around the weighted average of the DA for a given number of turns, i.e. $N_1=10^6$ (corresponding the the maximum number of turns simulated with SixTrack), $N_2=10^7$ and $N_3=10^8$ (corresponding to two extrapolation times). For $N_1$, a comparison between the DA obtained from numerical simulations and that derived from the various models is carried out. The DA data from SixTrack are averaged over the realisations and the rms is used as associated error. For the DA values obtained from the models, for each LHC lattice realisation a model is fitted, the errors on the corresponding model parameters are used to evaluate the error associated with the DA estimate. Finally, all sixty DA values are averaged using the corresponding errors as weights and the rms is used as associated error. The numerical values, including the relative errors obtained by taking the minimum and maximum DA value for the ensemble of sixty values are listed in Table~\ref{tab:LHCmodels}. It is clear (both from Fig.~\ref{fig:LHCDAvstime} and Table~\ref{tab:LHCmodels}) that the error bars for Model~1 are larger than those of the other models, thus confirming a better precision of the extrapolated DA values for Model~2,~3, and~4. 
{\renewcommand{\arraystretch}{1.8}
\begin{table}[!htbp]
	\centering
	\caption{DA values for Configuration~B for Beam~1 as obtained from the numerical data or by using the four models discussed in this paper, which provide also DA values extrapolated beyond the simulated number of turns. The approach used to derive the errors reported in the table is described in the main text.}
	\begin{tabular}{@{}l@{}@{}c@{}@{}c@{}@{}c@{}@{}c@{}@{}c@{}}
	\hline \hline
 Comment       & Turn number & Model 1 & Model 2 & Model 3 & Model 4 \\ \hline 
Numerical data        & $10^6$      & \multicolumn{4}{c}{$7.52 \pm 0.06^{+1.50\%}_{-1.53\%}$} \\ \hline
Models: interpolation & $10^6$      & $7.495 \pm 0.002^{+1.94\%}_{-1.60\%}$ & $7.572 \pm 0.003^{+2.11\%}_{-1.72\%}$ & $7.568 \pm 0.007^{+2.79\%}_{-2.24\%}$ &  $7.57 \pm 0.02^{+3.43\%}_{-2.89\%}$ \\ \cline{2-6}
\multirow{2}{35mm}{Models: extrapolation} 
                      & $10^7$      & $7.033 \pm 0.003^{+3.51\%}_{-2.78\%}$ & $7.292 \pm 0.003^{+2.34\%}_{-1.59\%}$ & $7.286 \pm 0.007^{+3.02\%}_{-2.12\%}$ & $7.29 \pm 0.02^{+3.64\%}_{-2.77\%}$ \\ \cline{2-6}
                      & $10^8$      & $6.575 \pm 0.003^{+5.77\%}_{-6.37\%}$  & $7.058 \pm 0.004^{+2.54\%}_{-1.50\%}$ & $7.048 \pm 0.007^{+3.21\%}_{-2.02\%}$ & $7.06 \pm 0.02^{+3.82\%}_{-2.66\%}$ \\ 
\hline \hline
	\end{tabular}
	\label{tab:LHCmodels}
\end{table}}

\begin{figure}[!htbp]
\centering
\begin{tabular}{@{}c@{}c@{}c@{}}
	\includegraphics[trim = 0.5mm 1mm 1mm 2.3mm, width=.33\linewidth,clip=]{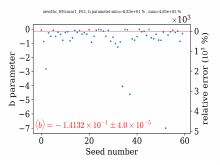} &
	\includegraphics[trim = 0.5mm 1mm 1mm 2.3mm, width=.33\linewidth,clip=]{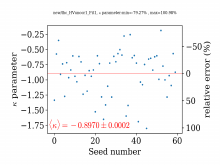} &
	\includegraphics[trim = 0.5mm 1mm 1mm 2.3mm, width=.33\linewidth,clip=]{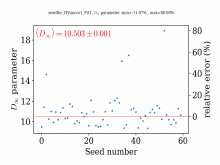} 
\end{tabular}
\caption{Distribution of the parameters of Model~1 for the sixty realisations corresponding to Configuration~B for Beam~1. The parameter value as well as its relative value (with respect to the average over the seed) is provided. The huge spread is clearly visible.}
\label{fig:LHCDAvstime_fit1}
\end{figure}

Another essential feature is that Model~1 provides parameters that vary significantly between different realisations of the magnetic errors (called seeds), also changing their signs. This is clearly shown in Fig.~\ref{fig:LHCDAvstime_fit1}, where the three parameters of Model~1 are shown as a function of the seed. The average value of each parameter over the seeds is also given, weighted by the corresponding error from the fitting procedure, together with an estimate of the associated error. As additional information, the relative spread around the weighted average of each model parameter is shown using the secondary vertical axis of each plot, and the wide range covered can be clearly appreciated. 

\begin{figure}[htb]
\centering
\begin{tabular}{@{}c@{}}
{\includegraphics[trim = 12mm 10mm 10mm 25mm, width=0.5\linewidth,clip=]{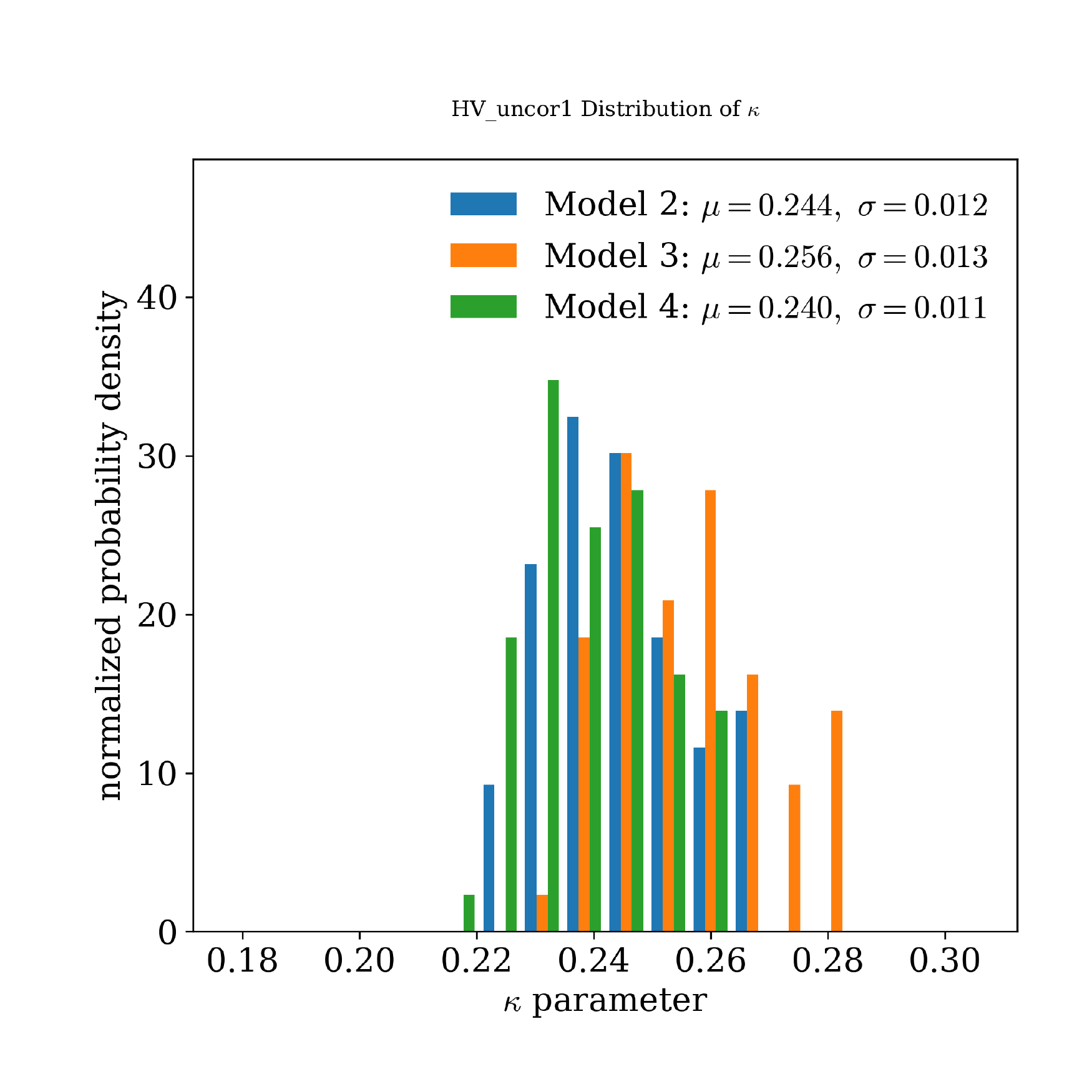}} \\
{\includegraphics[trim = 12mm 10mm 10mm 25mm, width=0.5\linewidth,clip=]{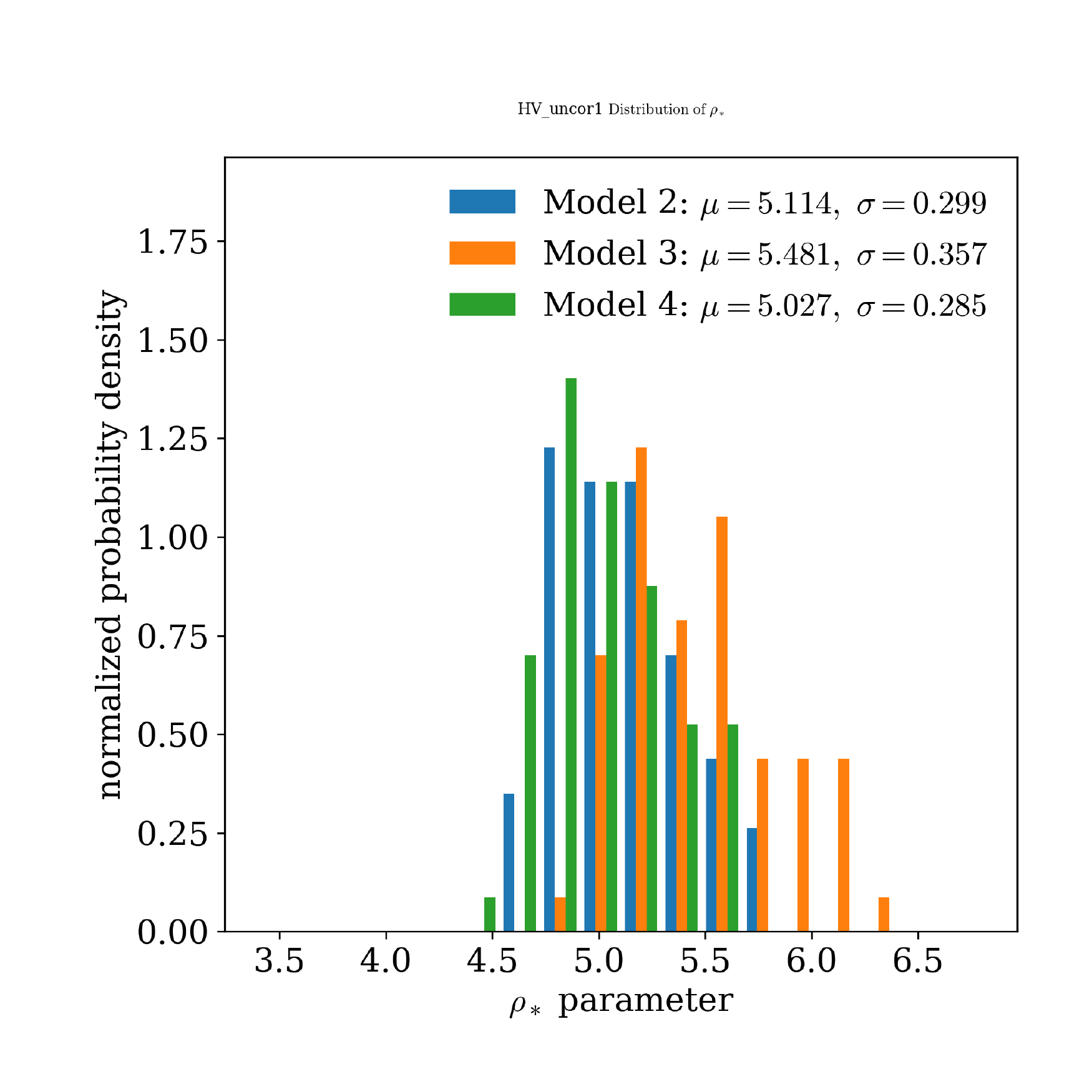}}
\end{tabular}
\caption{Distribution of the parameters for Model~2,~3, and~4 for the sixty realisations corresponding to Configuration~B for Beam~1. The distributions' parameters (average $\mu$ and $\sigma$) are also reported in the plots.}
\label{fig:LHCDAvstime_fitall}
\end{figure}

For the sake of comparison, the distributions of parameters for Model~2,~3, and~4 are reported in Fig.~\ref{fig:LHCDAvstime_fitall}. The situation is completely different, with only positive values and a rather small spread between the different cases, hence overcoming the limitations observed for the original Model~1. 

For some cases, even Model~1 provides positive values of the parameters, as observed already in~\cite{invlog}. However, whenever Model~1 provides negative nonphysical model parameters, the new models provide positive values, without meaningful impact on the accuracy in reproducing the numerical data. Indeed, the main conclusion of our investigations is that the very form of the original model, with the constant term $D_\infty$ added to the logarithmic one, is responsible for the appearance  of negative nonphysical model parameters. Furthermore, it induces a larger variability of the model parameters, possibly indicating overfitting. All these aspects are absent in the newly proposed models for the DA dependence on time, which is an important step forward for reliable modelling of the time dependence of DA. 

\section{\protect  Probing the predictive power of the DA models} \label{sec:predictivity}
{ As it was presented and discussed in~\cite{invlog}, the predicting power of the proposed DA models has been verified by varying the amount of data used to build the models and then extrapolating the DA value to a fixed number of turns to verify the agreement between the DA value obtained by means of the models and the numerical simulations. Indeed, in the previous section the extrapolation properties up to $10^8$ turns have been studied, but without comparing against tracking data, which is nowadays still impossible for such a large number of turns in the case of the LHC ring. Therefore, in this section a different aspect is considered, which consists of benchmarking the performance of the extrapolation against numerical data by choosing a suitable number of turns that can be simulated without too many issues. 

In Fig.~\ref{fig:Henon_extra}, the results of the DA extrapolation obtained from the proposed models built using different number of turns for the H\'enon map are shown. Model~2, 3, and 4 are reported as a function of the parameter $\epsilon$ and the error bar are obtained from the fit procedure. The numerical simulations carried out at $10^7$ turns are considered the reference and are reported without any error bar as no fitting and extrapolation are involved. 
\begin{figure}[htb]
\centering
{\includegraphics[trim = 11mm 2mm 20mm 12mm, width=0.7\linewidth,clip=]{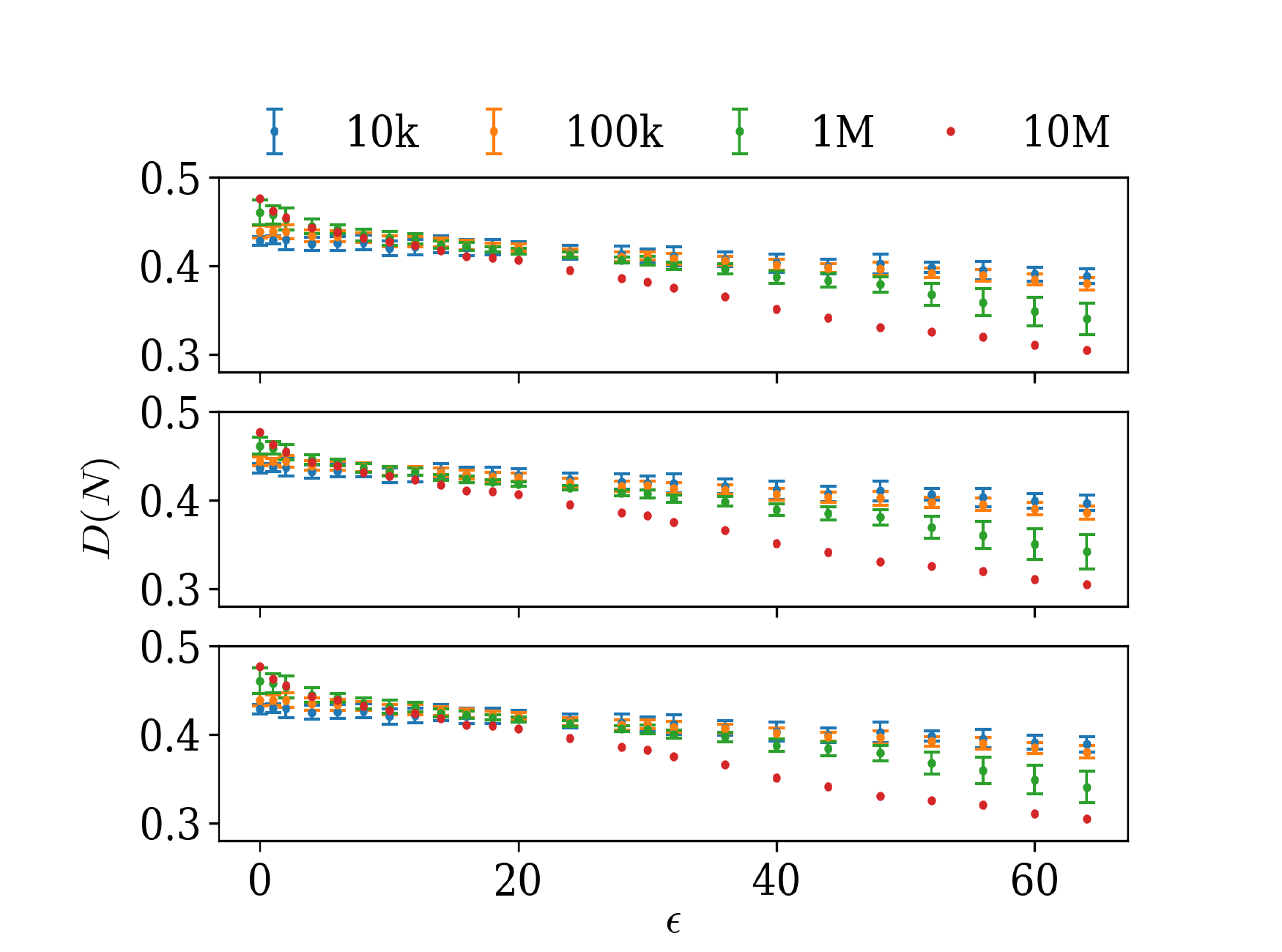}}
\caption{ Results of the DA extrapolation obtained from the proposed models built using different number of turns (marked on the upper part of the plot) for the H\'enon map. The cases of Model~2, 3, and 4 are shown here (top to bottom), all as a function of the parameter $\epsilon$. The error bars are obtained from the fit procedure. The reference values are those obtained from numerical simulations performed with $10^8$ turns, which are shown without error bars.}
\label{fig:Henon_extra}
\end{figure}

The first observation is that the overall behaviour does not depend on the DA model used. For $\epsilon \leq 20$, the extrapolated DA values approximate the reference ones from below and the agreement is very good and weakly depends on the number of turns used to build the DA models. In several cases the reference DA values are compatible with the extrapolated ones within the error bars. Note that the relative agreement between reference and extrapolated DA is better than $\approx 5$\%. The conclusions change whenever the range $\epsilon \geq 20$ is considered. In fact, a stronger dependence on $\epsilon$ is observed and the extrapolated DA values approach the reference ones from above. A larger discrepancy between extrapolated values and references ones is observed, the maximum reaching $\approx 30$\% for the largest values of $\epsilon$. In any case, the extrapolation based on $10^6$ turns never differs by more than $\approx 15$\% from the reference. This can be considered as a very encouraging result for a time extrapolation by two orders of magnitude.

Similar analyses have been carried out for the LHC cases, applying a similar approach as that used for the H\'enon map. The results are summarised in Fig.~\ref{fig:LHC_extra} where the proposed models (2, 3, and 4) are presented (top to bottom). In the two columns the results obtained for different amount of numerical data ($10^4$ - left - $10^5$ - right) are shown. The outcome of the analyses for all the sixty realisations of the LHC ring are reported and the behaviour of the numerical data used to build the DA models is shown together with the extrapolated curves. The reference case is provided by the results of numerical simulations up to $10^6$ turns. 

\begin{figure}[p]
\begin{center}
\begin{tabular}{@{}c@{}@{}c@{}}
{\includegraphics[trim = 16mm 8mm 21mm 24mm, width=0.49\linewidth,clip=]{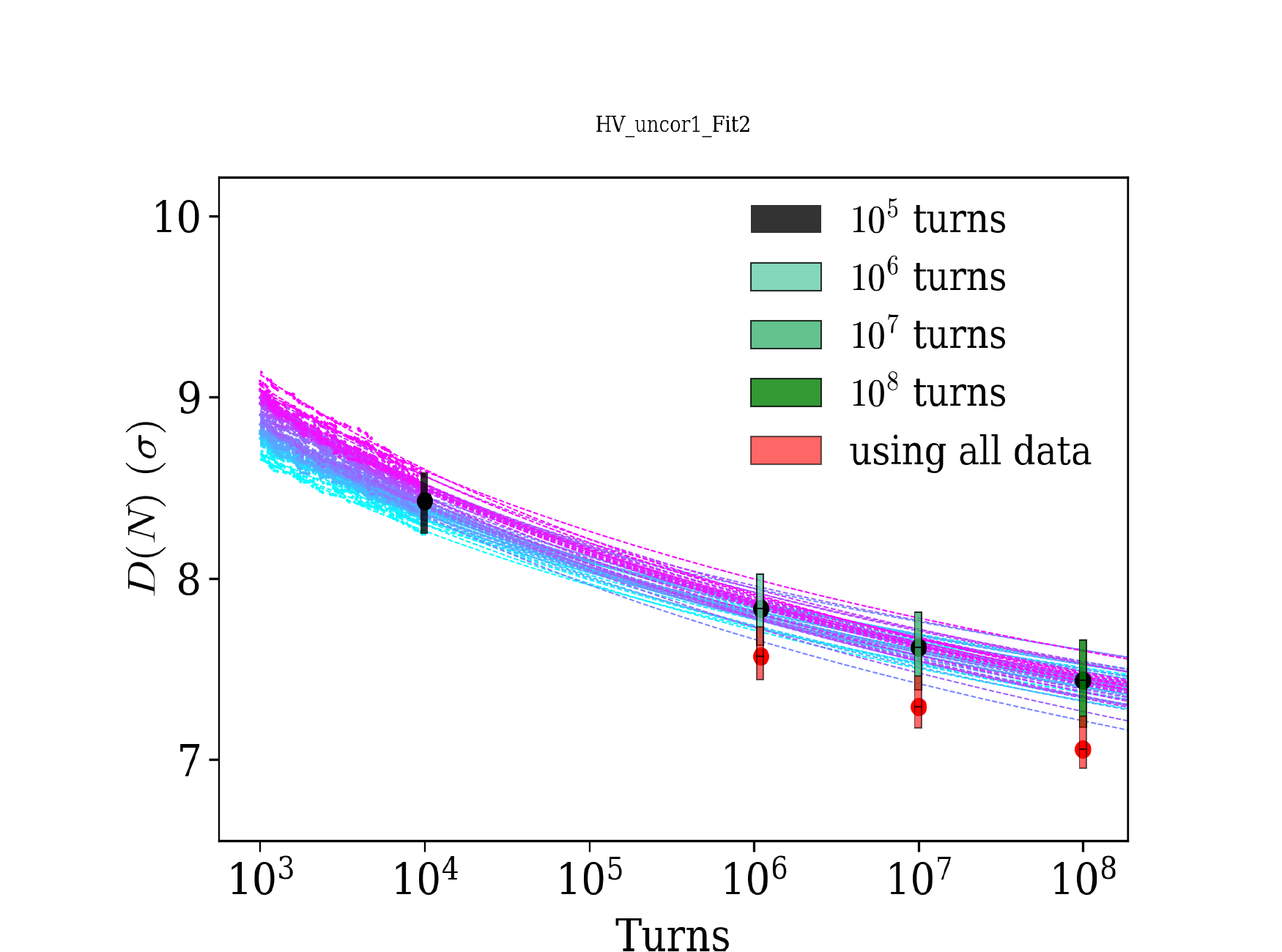}} & 
{\includegraphics[trim = 16mm 8mm 21mm 24mm, width=0.49\linewidth,clip=]{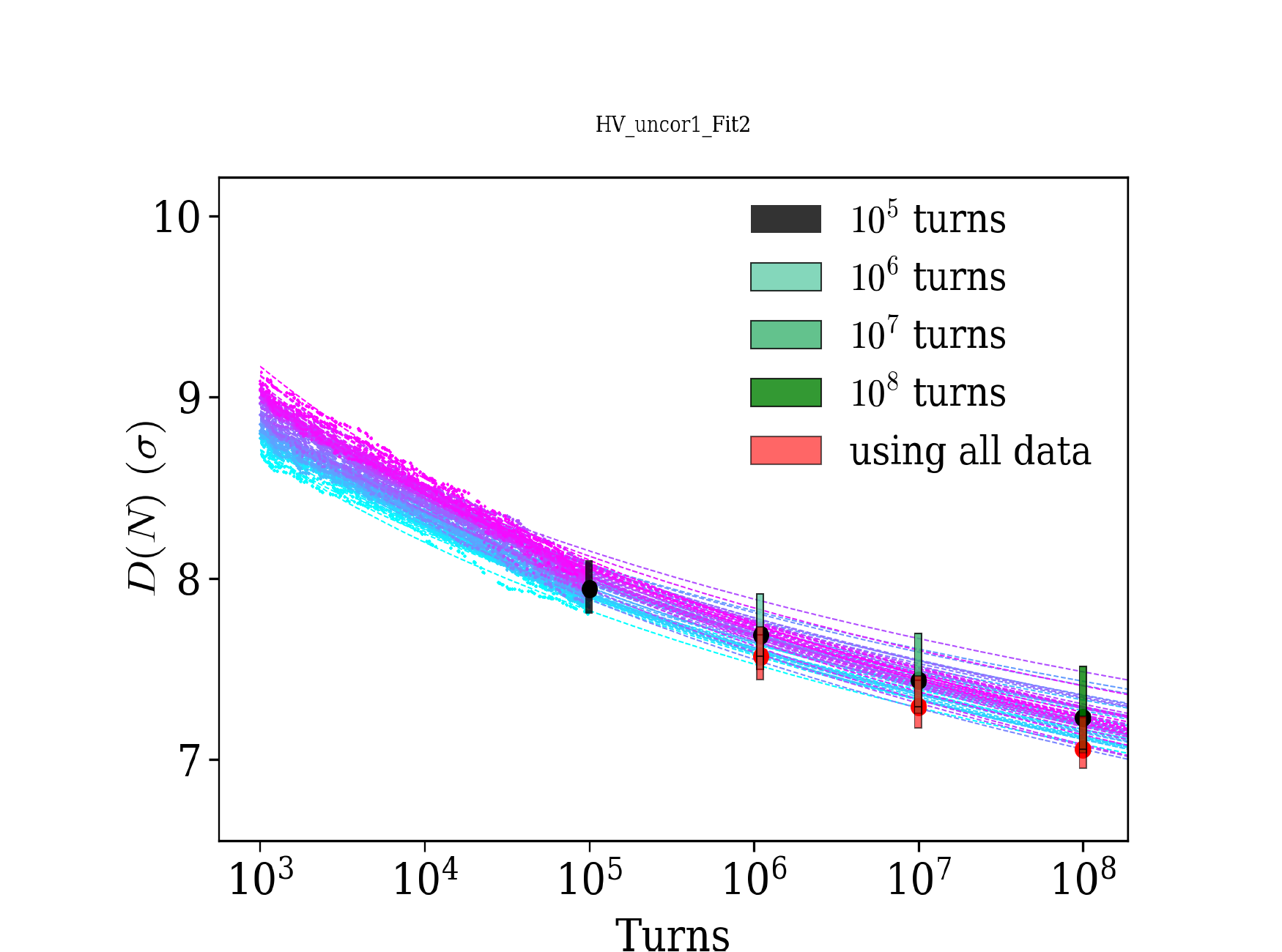}} \\
{\includegraphics[trim = 16mm 8mm 21mm 24mm, width=0.49\linewidth,clip=]{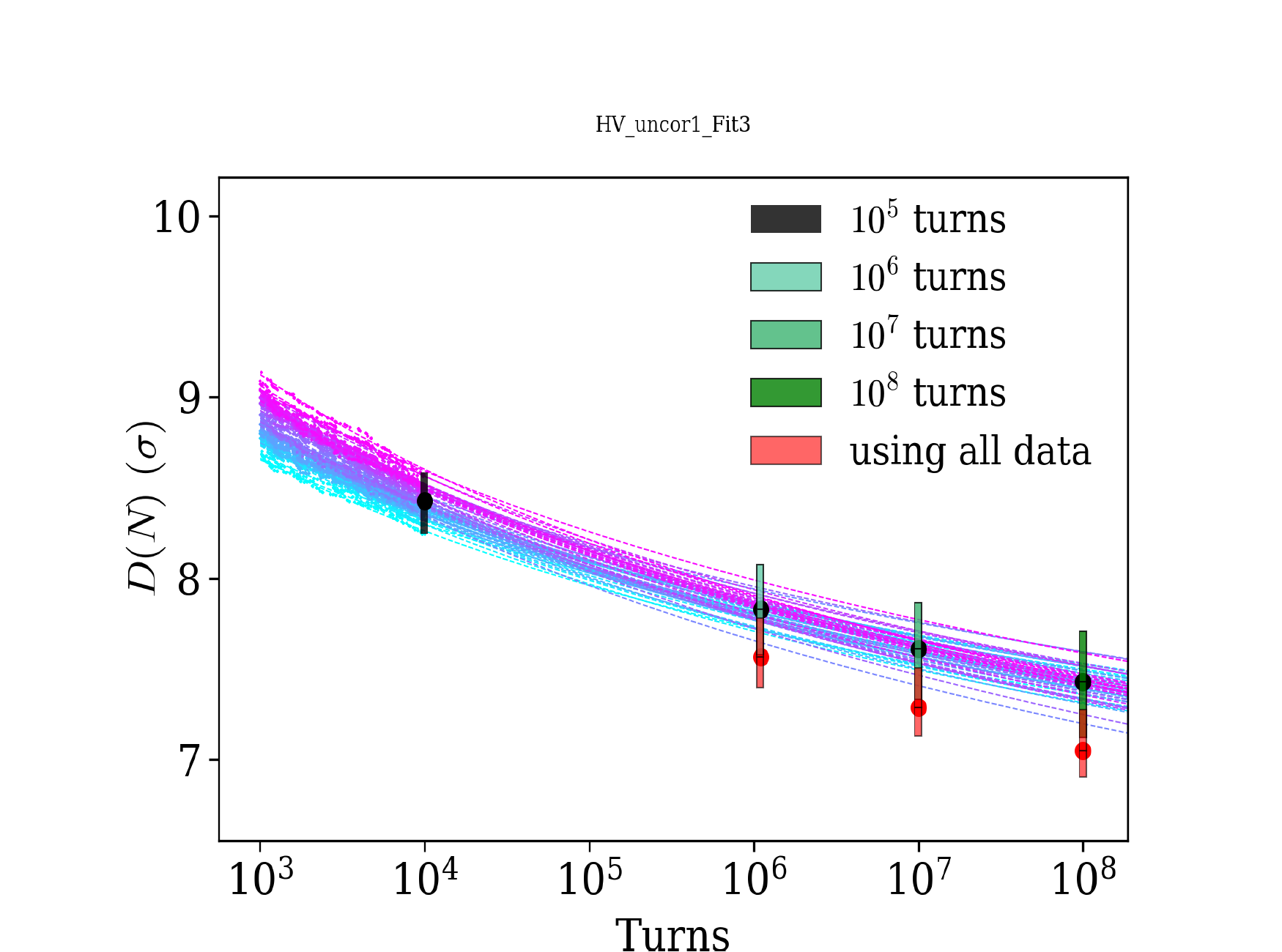}} & 
{\includegraphics[trim = 16mm 8mm 21mm 24mm, width=0.49\linewidth,clip=]{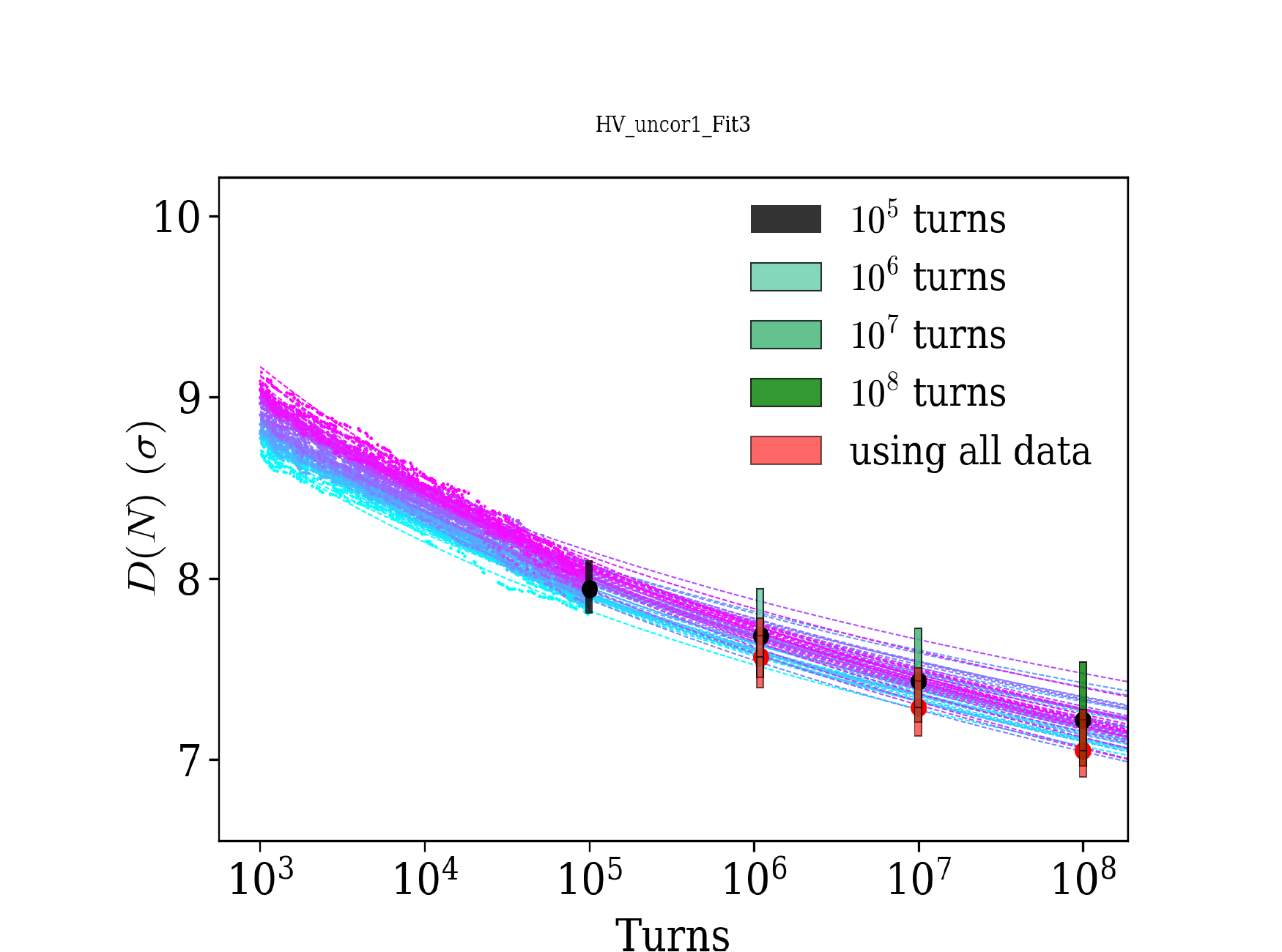}} \\
{\includegraphics[trim = 16mm 0mm 21mm 24mm, width=0.49\linewidth,clip=]{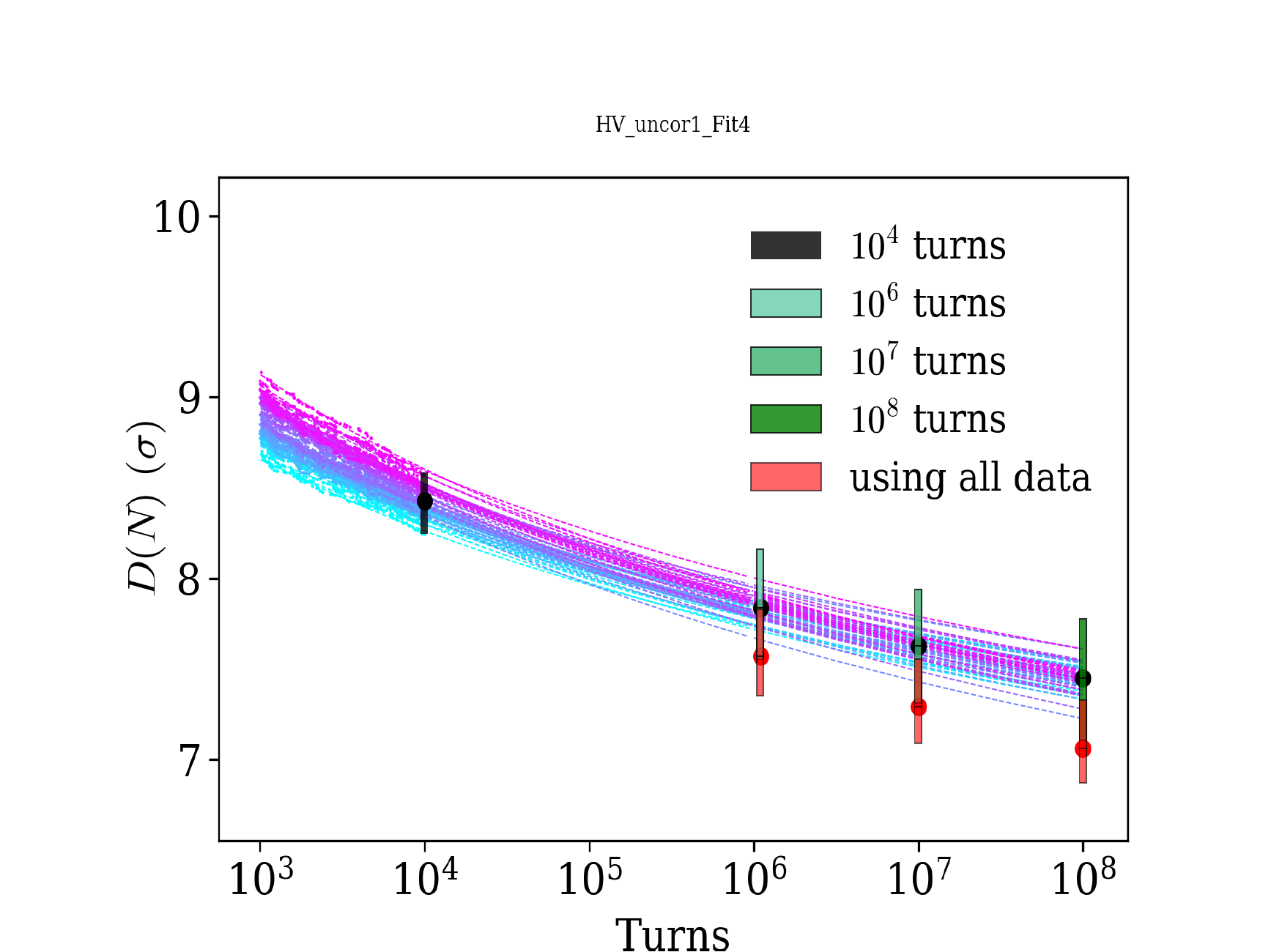}} & 
{\includegraphics[trim = 16mm 0mm 21mm 24mm, width=0.49\linewidth,clip=]{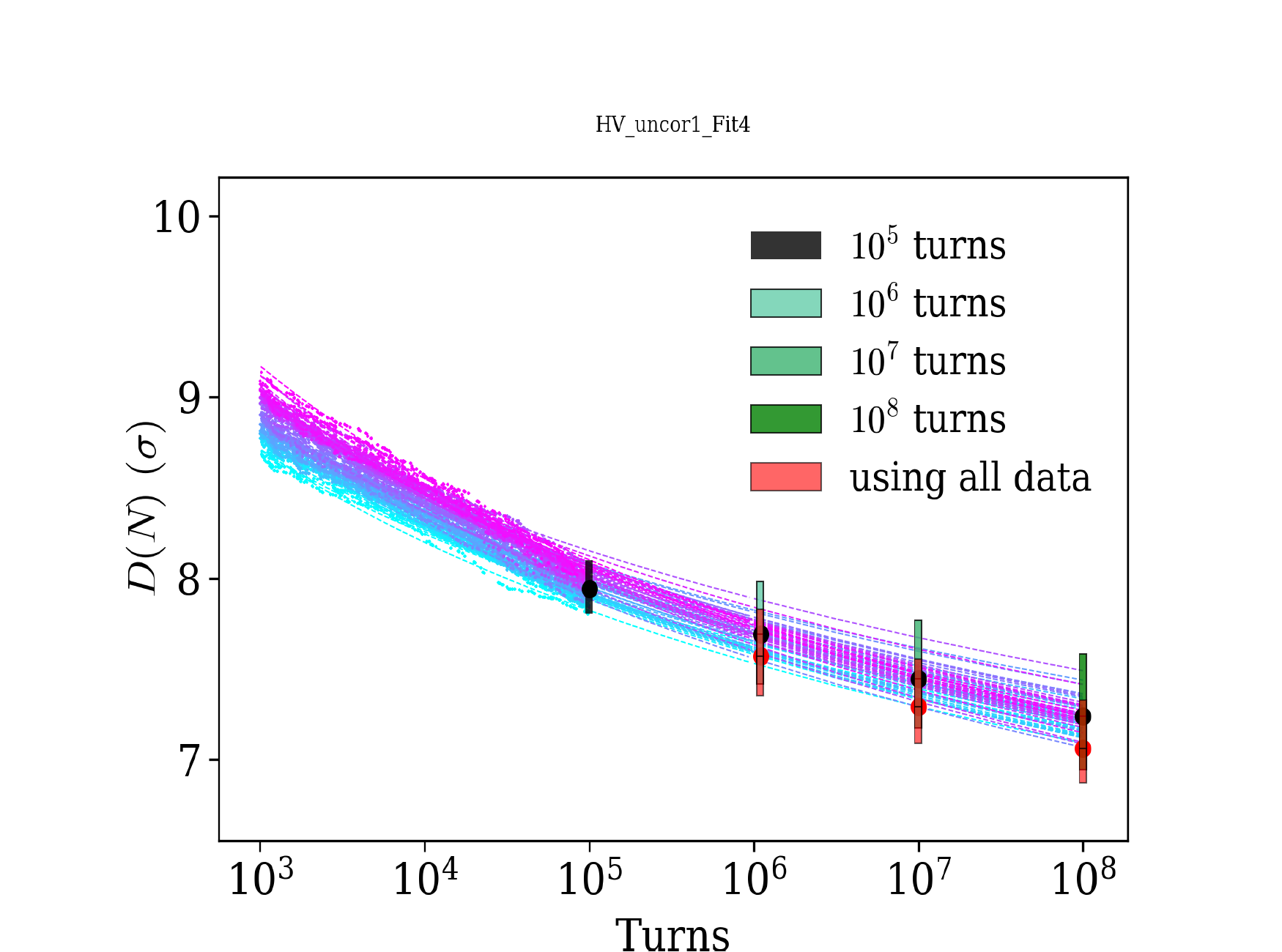}} 
\end{tabular}
\caption{ Results of the DA extrapolation obtained from the proposed models built using different number of turns ($10^4$ - left - $10^5$ - right) for the LHC case corresponding to Beam~1 with Configuration~B. The cases of Model~2, 3, and 4 are shown here (top to bottom). The results for all sixty realisations of the LHC ring are plotted, with the numerical data used to build the DA models shown together with the extrapolated curves up to $10^8$ turns.}
\label{fig:LHC_extra}
\end{center}
\end{figure}

The results presented in both columns approximate the reference data from above, showing that the extrapolated DA is an optimistic estimate of the numerical one. As expected, the larger is the number of turns used to build the DA models the smaller is the discrepancy between extrapolated and reference values. The quantitative data are reported in Table~\ref{tab:LHCmodels_ext}.
{\renewcommand{\arraystretch}{1.8}
\begin{table}[!htb]
	\centering
	\caption{ DA values for Configuration~B for Beam~1 as obtained from the numerical data or by using the three new models discussed in this paper, which provide DA values extrapolated to $10^6$ turns, but starting from a different number of turns used to build the models. The approach used to derive the errors reported in the table is described in the main text.}
	\begin{tabular}{@{}l@{}@{}cccc@{}}
	\hline \hline
 Comment       & Turn number  & Model 2 & Model 3 & Model 4 \\ \hline 
Numerical data & $10^6$      & \multicolumn{3}{c}{$7.52 \pm 0.06^{+1.50\%}_{-1.53\%}$} \\ \hline
\multirow{2}{45mm}{Models: extrapolation from} 
                      & $10^4$      & $7.687 \pm 0.004^{+2.94\%}_{-2.50\%}$ & $7.83 \pm 0.01^{+3.16\%}_{-3.18\%}$ & $7.83 \pm 0.02^{+4.13\%}_{-3.63\%}$ \\ \cline{2-5}
                      & $10^5$      & $7.687 \pm 0.004^{+2.94\%}_{-2.50\%}$ & $7.684 \pm 0.009^{+3.37\%}_{-3.02\%}$ & $7.69 \pm 0.02^{+3.80\%}_{-3.55\%}$ \\ 
\hline \hline
	\end{tabular}
	\label{tab:LHCmodels_ext}
\end{table}}

The disagreement between extrapolated DA values and the reference ones ranges from $2$\% to $4$\%, which is an excellent result. No strong dependence of the extrapolated DA value on the model used has been observed as for the case of the H\'enon map. As a consequence, extrapolation for two orders of magnitude in turn number might be at hand also for the LHC. Of course, it is not possible to ensure that such a situation is typical for all possible configurations of the LHC (similarly to what has been observed for the H\'enon map, where the degree of accuracy of the extrapolation features a non-negligible dependence on the value of the parameter $\epsilon$.}
\section{Digression: some intriguing properties of the DA models} \label{sec:digression}
\begin{figure}[htb]
\begin{center}
\begin{tabular}{@{}c@{}}
{\includegraphics[trim = -7mm 0mm 0mm 0mm, width=0.5\linewidth,clip=]{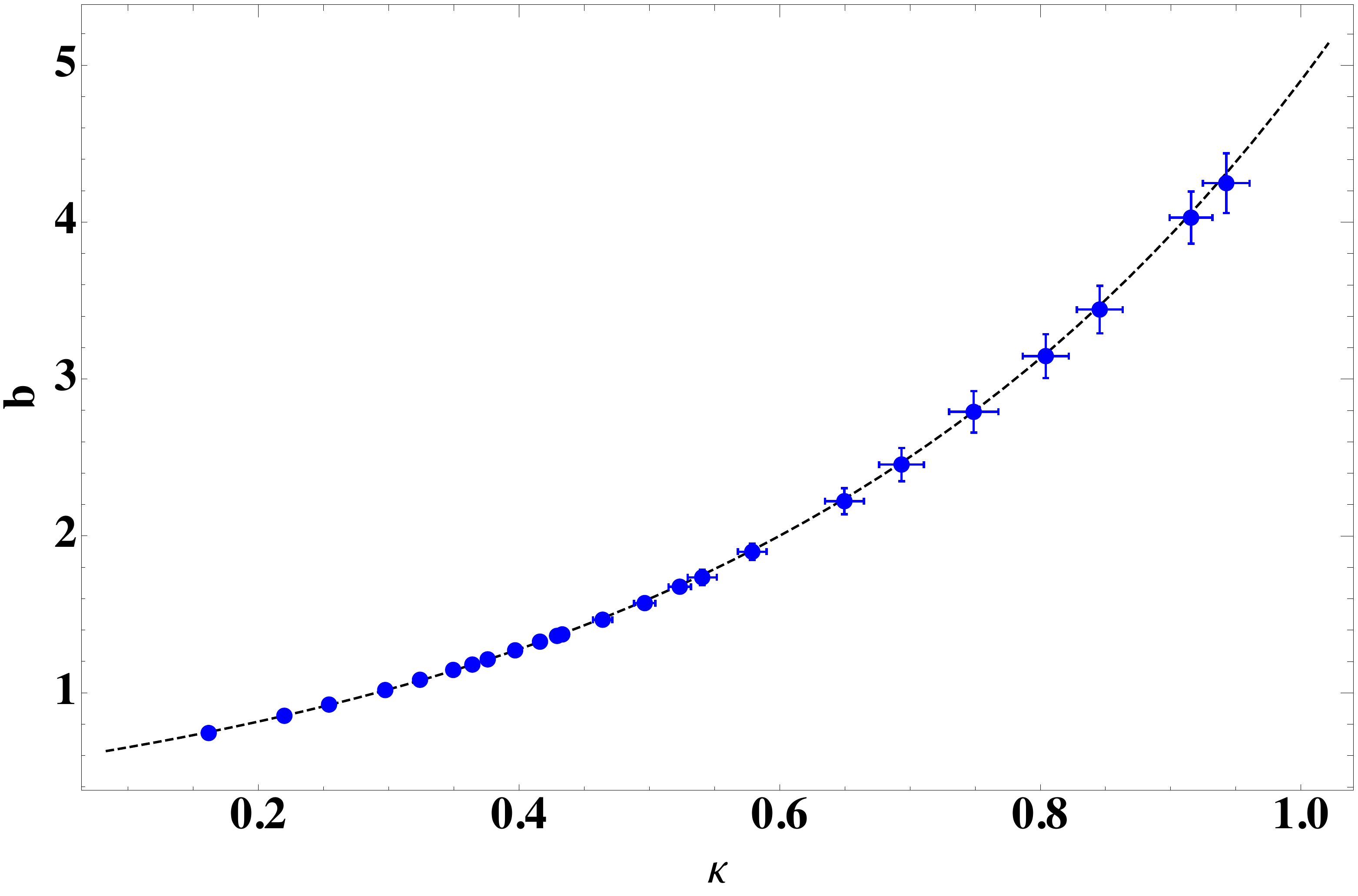}} \\
{\includegraphics[trim = 0mm 0mm 0mm 0mm, width=0.5\linewidth,clip=]{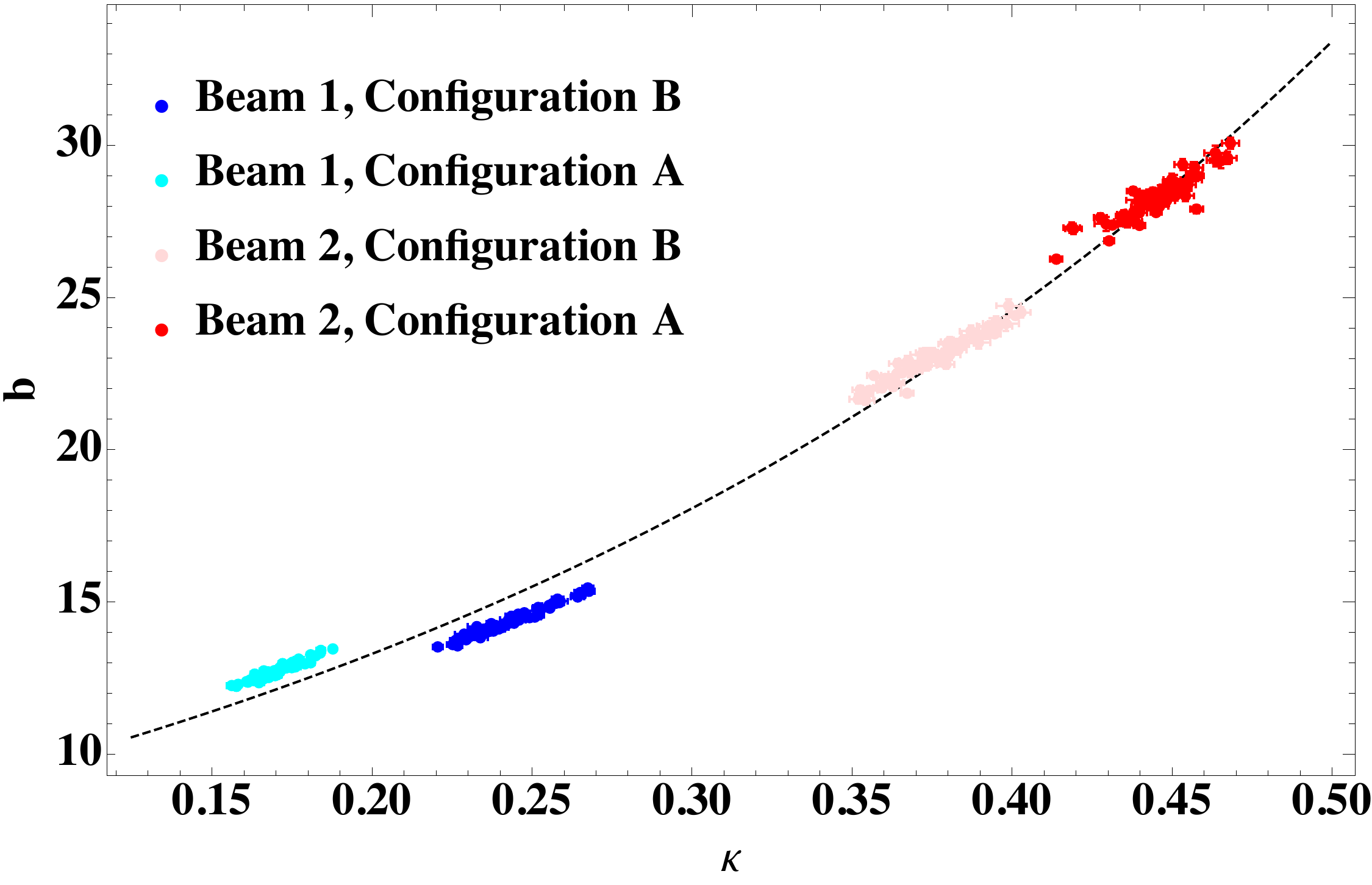}}
\end{tabular}
\caption{Dependence of $b$ vs $\kappa$ for the H\'enon map (upper) and the four LHC cases (lower) considered earlier. The error bars associated with the fit procedure used to determine $b$ and $\kappa$ are also shown. The data shown refer to Model~2, but similar results hold also for Model~3 and~4. The dashed curves represent the fit function~\eqref{eq:mod1}.}
\label{fig:digression}
\end{center}
\end{figure}

The DA models provide values of the parameters $\rho_\ast$ or $b$ (see Eqs.~\eqref{model2.1_1} and~\eqref{bvskappa}) and $\kappa$. It is of interest to verify whether the assumed relationship~\eqref{parameter2} is confirmed by the data. To this aim, the dependence of $b$ on $\kappa$ has been studied. The results for Model~2 are shown in Fig.~\ref{fig:digression}, where the H\'enon map and LHC models are reported in the upper and lower part of the plot, respectively.

The striking observation is that, in spite of the essential differences in the four LHC rings' configurations, all data seem to lie on a single smooth curve. The same holds also for the H\'enon map case, where the differences in the value of $\epsilon$ do not prevent the data points to lie on a single smooth curve. Furthermore, the functional form of the curve is the same for the H\'enon and LHC cases, showing a sort of universal behaviour. Note also that these features are model independent, as they hold also for Model~3 and~4. 

{ It is worthwhile making two additional remarks. Firstly, the difference between Beam~1 and~2 in terms of the parameters $b$ and $\kappa$ is not unexpected. In fact, it was already observed in earlier work~\cite{DAasbuilt} that the DA for $10^5$ turns was not the same for the two rings. This is linked to the fact that the magnetic field errors are rather correlated between the two ring, but they are not exactly the same. Secondly, the parameter $\kappa$ varies although the phase space dimension is not changing: this is yet another indication that further theoretical work is needed to understand the dependence of $\kappa$ on the features of the system.}

Several functions have been considered to describe the observed dependence, namely
\begin{subalign}
b(\kappa) & = \alpha \exp{\beta \,\kappa} \label{eq:mod1}\\
b(\kappa) & = \alpha \exp{\beta \,\sqrt{\kappa}} \label{eq:mod2}\\
b(\kappa) & = \alpha \exp{\beta \,\kappa^\gamma} \label{eq:mod3}\\
b(\kappa) & = \alpha \, (\beta \, \kappa)^{\gamma \, \kappa} \label{eq:mod4}
\end{subalign}
with $R^2_{\rm adj}$ as figure of merit. 

All listed functions provide reasonably good fit results, with model~\eqref{eq:mod3} generally giving the best fit and model~\eqref{eq:mod1} being a close second. However, for model~\eqref{eq:mod3} we always obtain  $\gamma\approx 1$, hinting that model~\eqref{eq:mod1} might be the correct underlying relation. The most important observation, however, is that the function~\eqref{eq:mod4} always provided the worst fit, despite it being expected to be the correct underlying relation (see Eq.~\eqref{bvskappa}). This led us to conjecture that the  function~\eqref{eq:mod1} represents the correct underlying relation between $b$ and $\kappa$.

It is worth noting that tests adding constant terms to the listed fit functions were carried out. Once more, it was found that although the constant term helps in improving the fit quality, it makes the values of the other fit parameters entering in the $\kappa$-dependent term more variable when applied to the several systems under consideration. For this reason, it has been considered that such a constant term introduces nonphysical features and it has been dropped. The resulting fit parameters for model~\eqref{eq:mod1} for the various configurations studied are listed in Table~\ref{tab:digression1}. 
{\renewcommand{\arraystretch}{1.3}
\begin{table}[!htbp]
	\centering
	\caption{Parameters of the fit function~\eqref{eq:mod1} describing the relationship between $b$ and $\kappa$, for the H\'enon map and the various LHC configurations, for Model~2. The errors are the standard errors associated with the fit procedure. Moreover, the errors associated with the determination of $b$ and $\kappa$ have been used as weights for determining the fit function~\eqref{eq:mod1}.}
	\begin{tabular}{cl@{}|ccc}
	\hline \hline
&                 & \multicolumn{3}{c}{Model~2} \\
& Configuration   & & & \\
&                 & $\alpha$ & $\beta$ & $R^2_{\rm adj}$ \\
	\hline
& H\'enon map                   & $0.524 \pm 0.001$ & $2.225 \pm 0.004$ & $99.999$ \\ \hline
\multirow{2}{10mm}{LHC\\Beam~1} 
&Configuration~A  & $7.5 \pm 0.2$ & $3.1    \pm 0.1$ & $99.996$ \\ \cline{2-5}
&Configuration~B  & $7.3 \pm 0.1$ & $2.80   \pm 0.07$ & $99.996$ \\ \hline
\multirow{2}{10mm}{LHC\\Beam~2}
&Configuration~A  & $11.4 \pm 0.6$ & $2.0   \pm 0.1$ & $99.989$ \\ \cline{2-5}
&Configuration~B  & $9.5 \pm 0.3$ & $2.35   \pm 0.08$ & $99.992$ \\ \hline \hline 
& All LHC configs.       & $7.18 \pm 0.06$ & $3.07 \pm 0.02$ & $99.996$ \\ \hline \hline
	\end{tabular}
	\label{tab:digression1}
\end{table}}

The error bars on the fit parameters for the four LHC cases are in general larger than those for the H\'enon map, which is a consequence of the presence of the sixty realisations for each LHC configuration. It is also clear that the parameters for each beam are relatively close together, in spite of the differences in magnetic configurations.  In general, the fit parameters for Model~3 and~4 are very similar between them, while those for Model~2 are more different. This is a possible indication that the observed scaling of $b$ with $\kappa$ is better detected by means of the more accurate models. Another observation is that the fit parameters of the combined data for all four LHC configurations do not depend strongly on the model used to describe the DA data.

We can now use Eq.~\eqref{eq:mod1} to redefine the proposed DA models such that the different fit parameters are supposed to be independent of each other. For convenience, we rename the fit parameters from model~\eqref{eq:mod1} as follows:
\begin{equation}
    \tilde{r}_* = \tilde{b} = \alpha \, ,
    \qquad\qquad
    \mathcal{B} = \exp{-\beta} \, ,
\end{equation}
such that we can rewrite \eqref{eq:mod1} as
\begin{equation}
    \tilde{r}_* =
        r_* \; \mathcal{B}^\kappa \, ,
    \qquad\text{or equivalently}\qquad
    \tilde{b} =
        b \; \mathcal{B}^\kappa \, .
\end{equation}
Note that, due to the dependency of $r_*$ on $\kappa$, both $\tilde{r}_*$ and $\mathcal{B}$ are independent on it. Furthermore, we define
\begin{equation}
    \tilde{N}_0 = N_0\, \mathcal{B}^{\lambda\kappa} \, ,
\end{equation}
which can be computed, using Eq.~\eqref{parameter3}, to be
\begin{equation}\label{NtildeEstimate}
    \tilde{N}_0 =
        \frac{7\sqrt{6}}{48}\,
        \tilde{b}^\lambda \, .
\end{equation}
Finally, this gives the following reformulation for our new DA models:
\hspace*{-1em}
\begin{subalign}[eq:newmodels]
    &\textbf{Model~2}\qquad \Rightarrow \qquad D(N) =
        \frac{\tilde{b}}{\left(
            \mathcal{B}\,\ln \displaystyle{\frac{N}{\tilde{N}_0}}
        \right)^\kappa} \ec \\
        & \notag \\
    &\textbf{Model~3}\qquad \Rightarrow \qquad D(N) = \tilde{b} \times \notag \\
        & \times \frac{1}{\left[
            \mathcal{B}\,\ln \displaystyle{\frac{N}{\tilde{N}_0}}
            +\displaystyle{\frac{\kappa\mathcal{B}}{2}} \ln \left(
                \mathcal{B}\,\ln\frac{N}{\tilde{N}_0}
                +\frac{\kappa\mathcal{B}}{2}\ln\frac{\kappa\mathcal{B}}{2}
            \right)
        \right]^\kappa} \ec \\
        & \notag \\
    & \textbf{Model~4}\qquad \Rightarrow \qquad D(N) = \tilde{b} \times \notag \\
        & \times \frac{1}{\left[
            -\displaystyle{\frac{\kappa\mathcal{B}}{2}}\,
            \mathcal{W}_{-1} \left(
                -\frac{2}{\kappa\mathcal{B}}
                \left(\frac{N}{\tilde{N}_0}\right)^{-\frac{2}{\kappa}}
            \right)
        \right]^\kappa} \ec
\end{subalign}
where $\tilde{N}_0$ can either be left as a free model parameter, or can be fixed to the value given in Eq.~\eqref{NtildeEstimate} for Models~3 and 4, or to an arbitrary constant for Model~2.

Note that these new formulations of the DA models introduce an extra fitting parameter, namely $\mathcal{B}$, but are supposed to be even more stable from a physical viewpoint. In other words, the parameters $\tilde{b}$, $\mathcal{B}$, and $\kappa$ are expected to be totally independent of each other, hence true constants of the scaling law. Of course, this is so far merely an empirical observation. Some efforts should be devoted to the analysis of the form of the estimate of the stability time provided by Nekhoroshev theorem to determine whether the numerically-obtained scaling law can be justified with theoretical arguments. In other words, an interesting open question is whether a theoretical motivation can be found to rewrite the Nekhoroshev estimate in Eq.~\eqref{eq: Nekhoroshev} as
\begin{equation}
	\frac{N(r)}{\tilde{N}_0} =
		\left(\frac{r}{\tilde{r}_\ast}\right)^{\!\lambda}\exp{\frac{1}{\mathcal{B}}\left(\frac{\tilde{r}_\ast}{r}\right)^{\frac{1}{\kappa}}}
		\ec
\end{equation}
where $\tilde{r}_\ast$, $\mathcal{B}$, and $\kappa$ are independent constants, and $\tilde{N}_0$ is given by Eq.~\eqref{NtildeEstimate}. Another interesting question is whether an analytical estimate can be found for $\mathcal{B}$, as our data suggests it to be constant for all different systems we investigated. Indeed, Table~\ref{tab:digression1} seems to hint that
\begin{equation}
      0.05 \leq \mathcal{B} \leq 0.14 \ep
\end{equation}
As a last remark, we note that it is straightforward to retrieve the original DA model formulations from Eqs.~\eqref{eq:newmodels} by simply making the following substitutions:
\begin{equation}
    \tilde{b} \to b, \qquad \tilde{N}_0 \to N_0, \qquad \mathcal{B} \to 1 \ep
\end{equation}
\section{Conclusions} \label{sec:conc}
In this paper, recent progress in defining reliable models for the time dependence of the DA has been presented and discussed in detail. The essence of the novel approaches relies on the analytic estimates used in the proof of the Nekhoroshev theorem~\cite{Bazzani:1990aa,Turchetti:1990aa} and on the use of the Lambert function. Such a function is applied to invert the estimate of the stability time with an exact and closed-form expression. Three new models for the DA evolution with time have been proposed: Model~2 resembles the original Model~1~\cite{dynap1,invlog} for the inverse logarithmic term, but does not include the constant term representing the dynamic aperture for infinite time, which was introduced to take into account the region filled by KAM tori. Model~4 represents the exact model for DA evolution with time based on Nekhoroshev theorem, only, and using the Lambert function. Model~3 is derived from Model~4 by means of a series development of the Lambert function, which can provide an easier computational tool as it avoids the use of the exotic Lambert function. Moreover, it shows explicitly the resemblance and the new features of Model~3 with respect to Model~2. Of course, the validity of Model~3 relies on the validity of the expansion of the Lambert function, which remains a topic deserving further theoretical investigations.

In earlier work it had been observed that in some cases the parameters of the original Model~1, based on KAM theory and Nekhoroshev theorem, become negative~\cite{dynap1,invlog}. This represents a violation of the conditions of validity of the Nekhoroshev theorem, which then makes Model~1 only a phenomenological description of the special cases with negative parameters. The three new models proposed in this paper overcome this difficulty as they provide positive physical parameters values in situations in which Model~1 is failing to do so. Further, only a small reduction of fit quality is obtained, despite the reduction of the number of model parameters from three to two, which suggests a more fundamental scaling law has been found. 

The observed dependencies between the parameters of Model~1 have been reduced in the new models. Furthermore, the behaviour of the models parameters has been studied as a function of the modulation amplitude $\epsilon$, for the case of the 4D H\'enon map, and of the realisations of the magnetic field errors, for the case of the LHC, and a very smooth and regular behaviour has been observed. This feature opens the possibility to study the underlying mechanisms and how $b$ and $\kappa$ depend on the physical parameters of the system, rather than just considering the dynamic aperture. This could provide a more fundamental insight into the beam  dynamics and aid effective collider design based on $b$ and $\kappa$ rather than on DA computation at a fixed number of turns. 

A by-product of the detailed analysis carried out has been the observation that the $b$ parameter features a clear exponential dependence on the $\kappa$ parameter. This implies that the functional form of the stability-time estimate as given in~\cite{Bazzani:1990aa} could be reviewed. Additional theoretical efforts shall be devoted to this intriguing result. { Furthermore, the relationship between $\kappa$ and the phase space dimension or other dynamical features of the system should be studied in more detail to clarify the results presented and discussed in this paper.}

The main issue of the original model has been identified in the functional form proposed, namely with a constant term added to a logarithmic one that depends on two model parameters: this combination provides a flexible functional form that can match a large variety of DA data, however, compensations between the model parameters is possible and affects the stability and predictivity of the DA model.

As far as the possibility to use the models to interpolate numerical data and then to make predictions beyond the maximum number of simulated turns, Models~2,~3, and~4 proved to be very reliable and more precise than Model~1 for the cases presented in this paper. { More quantitatively, the predictive power has been probed for the dynamical systems considered in this study and it has been found out that extrapolations by two orders of magnitude in number of turns can be done with inaccuracy in the DA estimate not larger than $4$\% for the LHC case. This is certainly a very positive result that suggests that the use of these models for extrapolating DA beyond what is currently possible to compute by means of numerical simulations is indeed a viable option.}

Finally, it is worth stressing that these recent and very encouraging results will be used to refine the approaches proposed in earlier work to estimate beam losses~\cite{da_and_losses} and more recently to model the evolution of the collider's luminosity in the presence of burn off and losses due to dynamic aperture~\cite{LumiI,LumiII}.
\appendix 
\section{General properties of the \texorpdfstring{$\mathcal{W}$}{W} function}\label{app:LambertW}
If we want to invert Eq.~\eqref{eq: Nekhoroshev} in order to interpret $r$ as the dynamic aperture and express its evolution as a function of the number of turns, we are undoubtedly left with a result involving the Lambert $\mathcal{W}$ function, see, \eg~\cite{Knuth} and references therein for an overview on this function and its application in physics as well as~\cite{FZ} for a recent application to accelerator physics, so we devote this section to the investigation of this rather exotic, but nevertheless ubiquitous, function.

The Lambert $\mathcal{W}$ function, also called the $\Omega$ function or product logarithm, is in fact a set of functions defined as the branches of the inverse of the product exponential function, namely
\begin{equation}
	y = x \,\exp{x} \; \Leftrightarrow \; x = \mathcal{W}(y) \ep
\end{equation}
By analytic continuation, $\mathcal{W}(z)$ is well-defined, but multi-valued on the full complex plane, with a branch cut along the negative axis at $\left ] -\infty,-\frac{1}{\text{e}} \right ]$. Its defining equation for any $z\in\mathbb{C}$ is
\begin{equation}\label{eq: LambertWdefeq}
	z = \mathcal{W}(z) \, \exp{\mathcal{W}(z)} \ep
\end{equation}

By using the very definition~\eqref{eq: LambertWdefeq} it is possible to show that
\begin{subalign}
\exp{\pm n \mathcal{W}(z)} & = \left [ \frac{z^n}{\mathcal{W}^n(z)}\right ]^{\pm 1}  \\
\exp{\pm \mathcal{W}^n(z)} & = \left [ \frac{z}{\mathcal{W}(z)}\right ]^{\pm \mathcal{W}^{n-1}(z)} \ep
\label{eq: LambertWpropeq}
\end{subalign}

\begin{figure}[!t]
\centering
	\includegraphics[width=0.5\linewidth,clip=]{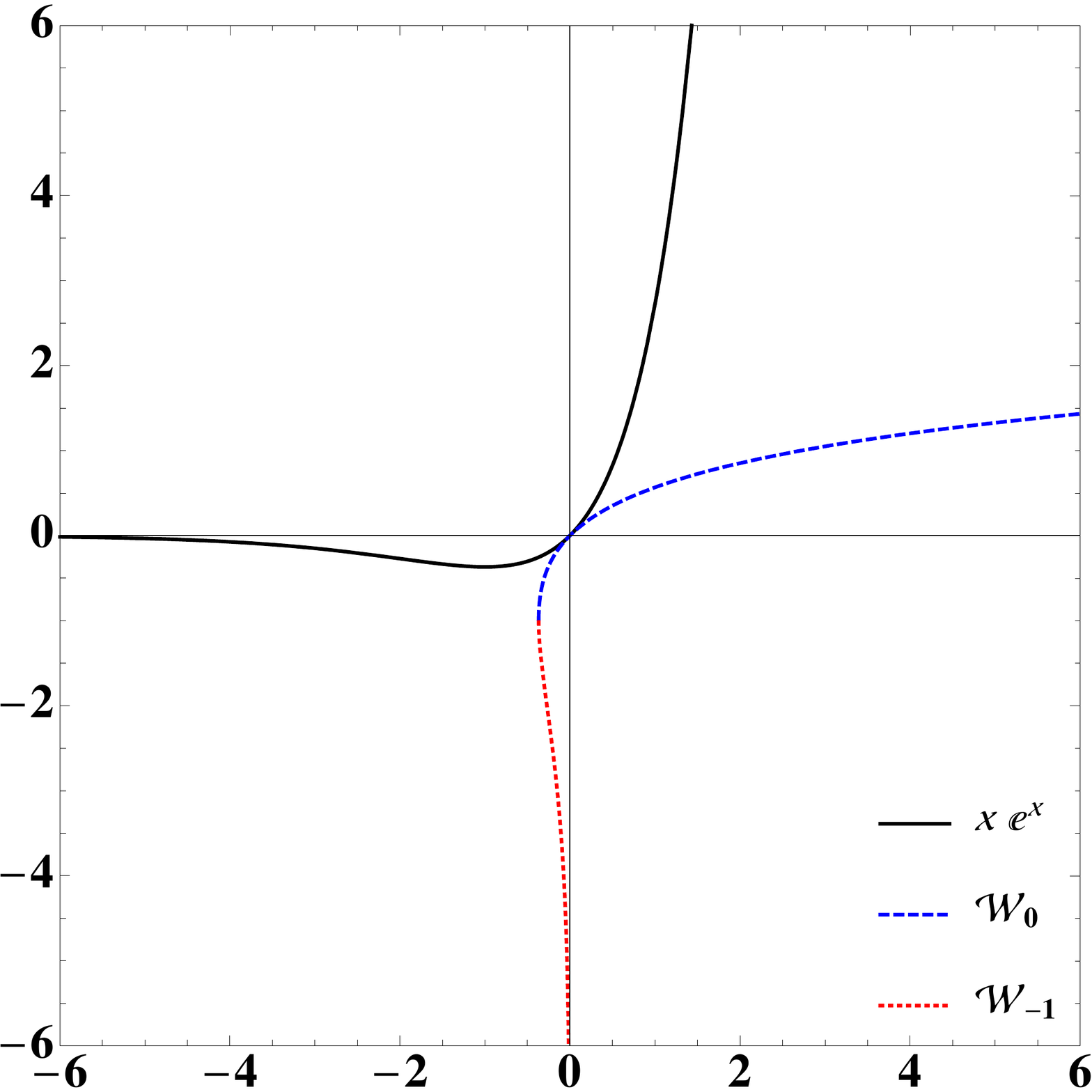}
	\caption{Plot of the product exponential function (full line) and the two real branches of its inverse, the Lambert $\mathcal{W}_0$ and $\mathcal{W}_1$ functions (dashed and dotted lines, respectively).}
	\label{fig: LambertW}
\end{figure}

Its derivative and primitive are the same for all branches and are given by
\begin{subalign}
	\Diff{\mathcal{W}(z)}{z} &=
		\frac{1}{z + \exp{\mathcal{W}(z)}} \ec \qquad \text{ for } z \neq -\frac{1}{\text{e}}\\
	\int\Dif{x}\mathcal{W}(x) &=
		x \mathcal{W}(x) -x + \text{e}^{\mathcal{W}(x)} + c \ep
\end{subalign}
Restricting ourselves to the real domain, there are only two branches of $\mathcal{W}$, namely
\begin{subalign}
	&\text{dom}\left(\mathcal{W}_0\right) =
		{\textstyle \left[-\frac{1}{\text{e}},+\infty\right[ }
		&\text{with}& -1 \leq \mathcal{W}_0 < + \infty \ec \\
	&\text{dom}\left(\mathcal{W}_{-1}\right) =
		{\textstyle \left[-\frac{1}{\text{e}},0\right[ }
		&\text{with}& -\infty \leq \mathcal{W}_{-1} \leq -1 \ec
\end{subalign}
with particular values
\begin{subalign}
	\mathcal{W}_0 (0) &= 0 \ec  &\mathcal{W}_{-1} (0^-) &= -\infty \ec \\
	{\textstyle \mathcal{W}_0 \left(-\frac{1}{\text{e}}\right)} &= -1 \ec &
	{\textstyle \mathcal{W}_{-1} \left(-\frac{1}{\text{e}}\right)} &= -1 \ep
\end{subalign}
These two branches are shown in Fig. \ref{fig: LambertW}.
A few other interesting identities, valid on the real axis, are:
\begin{subalign}
	\mathcal{W}_{0}\left(x \, \exp{x}\right) &= x \quad(x \geq -1) \ec \\
	\mathcal{W}_{-1}\left(x \, \exp{x}\right) &= x \quad(x \leq -1) \ec \\
	\mathcal{W}(x) &=
		\ln \frac{x}{\mathcal{W}(x)}
		\quad\left(x \geq {\textstyle -\frac{1}{\text{e}}}\right) \ec \\
	\mathcal{W}(x\,\ln x) &= \ln x \quad (x>0) \ec \\
		&= \mathcal{W}(x) +\ln \mathcal{W}(x) \quad (x>0) \ep
\end{subalign}
\section{Parameters' constraints}\label{app:parameters}
We investigate what constraints can be imposed on the parameters of our model. We have two general constraints: first we demand that the dynamic aperture is real and positive, and second we have to remain in the region where the Nekhoroshev estimate is valid, as in Eq.~\eqref{eq: NekhoroshevRegion}.
\subsection{Reality Condition}
For the first demand we have to investigate the argument of the Lambert $\mathcal{W}$ function, and the base of the power $-\kappa$. Indeed, if $\kappa$ is non-integer the result will be real only if the base of the power is positive. First of all, $r_\ast$ is a positive quantity, which ensures that $r>0$ as well as $N_0$. Furthermore if $\lambda\kappa <0$, the argument of $\mathcal{W}$ is positive and hence we use the upper branch $\mathcal{W}_0$, which is positive itself, thus ensuring the reality of the power, namely 
\begin{equation}\label{eq: RealityCondForLambdaKappaNeg}
	\lambda\kappa < 0 \quad \Rightarrow \quad
	r = r_\ast \left[
			|\lambda\kappa| \,\,\mathcal{W}_0\!\!\:\!\left(
					\frac{1}{|\lambda\kappa|}
					\sqrt[\leftroot{-3}\uproot{3}|\lambda\kappa|]{\frac{N}{N_0}}
			\right)
		\right]^{\!-\kappa}
		> 0 \ep
\end{equation}

On the other hand, if $\lambda\kappa > 0$, the argument of $\mathcal{W}$ is negative and hence we can use both branches, as both return a negative value for a negative argument, again ensuring the reality of the power. However, we have to make sure that the argument lies in the region $\left[-\frac{1}{\text{e}},0\right[$ to avoid entering a complex branch. In other words:
\begin{equation}\label{eq: RealityCondForLambdaKappaPos}
\begin{split}
	\lambda\kappa  > 0 \, \Rightarrow \,
	r & = r_\ast \left[
			-\lambda\kappa \,\,\mathcal{W}\!\!\:\!\left(
					-\frac{1}{\lambda\kappa}
					\sqrt[\leftroot{-3}\uproot{3}\lambda\kappa]{\frac{N_0}{N}}
			\right)
		\right]^{\!-\kappa}
		\!\!\!\! >0 \\
		\text{ if } &\,\,
	\frac{1}{\lambda\kappa} \!\!\sqrt[\leftroot{-3}\uproot{3}\lambda\kappa]{\frac{N_0}{N}}
	\leq \frac{1}{\text{e}}
	\ep
\end{split}
\end{equation}
\subsection{Region Constraint}
Next we have to satisfy Eq.~\eqref{eq: NekhoroshevRegion}. In other words, we have:
\begin{equation}
	\left[
			-\lambda\kappa \,\,\mathcal{W}\!\!\:\!\left(
					-\frac{1}{\lambda\kappa}
					\sqrt[\leftroot{-3}\uproot{3}\lambda\kappa]{\frac{N_0}{N}}
			\right)
		\right]^{\!-\kappa}
	\leq
	\left( \frac{2}{3\kappa} \right)^{\!\!\kappa}
	\ep
\end{equation}

To get rid of the power, we raise each side to $\frac{1}{\kappa}$. The result will hence depend on the sign of $\kappa$. Note that we already deduced that the base of the \lhs is positive. If $\kappa>0$ the base of the \rhs is positive as well and we can cancel the two powers without changing the ordering
\begin{equation}
	\label{eq: TempRegionCondForLambdaPos}
		\mathcal{W}\!\!\:\!\left(
					-\frac{1}{\lambda\kappa}
					\sqrt[\leftroot{-3}\uproot{3}\lambda\kappa]{\frac{N_0}{N}}
			\right)
		\leq
		-\frac{3}{2\, \lambda}
	 \quad \lambda > 0 \quad \kappa > 0 \ep 
\end{equation}

In order to continue our investigation of the restrictions on the parameters, we would like to apply the inverse of the Lambert $\mathcal{W}$ function, namely the product exponential. It is however not trivial if and how this function preserves a given ordering. If we have a look at Fig. \ref{fig: LambertW}, we see that $x\,\exp{x}$ rises monotonously towards $+\infty$ for $x \geq -1$, and falls monotonously from $0$ for $x \leq -1$. From this information we can derive the following properties:
\begin{subalign}[eq: ProductExpOrdering]
	\label{eq: ProductExpOrdering1}
	-1 \leq x \leq y \Rightarrow &\, x \exp{x} \leq y \exp{y}, \\
	\label{eq: ProductExpOrdering2}
	x \leq y \leq -1 \Rightarrow &\, x \exp{x} \geq y \exp{y}, \\
	\label{eq: ProductExpOrdering3}
	x \leq -1 \,\text{and}\, 0 \leq y \Rightarrow &\, x \exp{x} \leq y \exp{y}, \\
	x \leq -1 \, \text{and} -1 \leq y \leq 0 \Rightarrow &\, \text{information on ordering} \nonumber \\
	\label{eq: ProductExpOrdering4}
	\phantom{x \leq -1 \, \text{and} -1 \leq y \leq 0 \Rightarrow} &\, \text{is lost.}
\end{subalign}

If $\lambda>0$ and $\kappa>0$ then both Eqs. \eqref{eq: RealityCondForLambdaKappaPos} and \eqref{eq: TempRegionCondForLambdaPos} have to be satisfied. From the former we deduce that the argument of the Lambert $\mathcal{W}$ function is negative and hence the function itself is negative. To see which one of Eqs. \eqref{eq: ProductExpOrdering} applies, we investigate two positive regions of $\lambda$, namely
\begin{subalign}
	\label{eq: PosLambdaCond1}
	-\frac{3}{2\lambda} &\leq -1 \quad\Rightarrow\quad 0 < \lambda \leq \frac{3}{2} \ec \\
	\label{eq: PosLambdaCond2}
	-\frac{3}{2\lambda} &> -1 \quad\Rightarrow\quad \lambda > \frac{3}{2} \ep
\end{subalign}
Note that the combination of \eqref{eq: PosLambdaCond1} and \eqref{eq: TempRegionCondForLambdaPos} implies automatically that we have to choose the branch $\mathcal{W}_{-1}$, as this is the only (real) branch that has values smaller than $-1$. Let us again investigate the two cases separately.

\vspace*{1.5\baselineskip}
\noindent {$\boxed{0 < \lambda \leq 3/2}$}
\vspace*{0.5\baselineskip}

This implies that now both the \lhs and the \rhs of \eqref{eq: TempRegionCondForLambdaPos} are smaller than $-1$, and hence ordering \eqref{eq: ProductExpOrdering2} applies:
\begin{equation}
	-\frac{1}{\lambda\kappa} \sqrt[\leftroot{-3}\uproot{3}\lambda\kappa]{\frac{N_0}{N}}
	\geq
	- \frac{3}{2\lambda} \exp{- \frac{3}{2\lambda}}
	\quad
	0 < \lambda \leq \frac{3}{2}, \quad \kappa > 0 \ep
\end{equation}

Multiplying both sides with a factor $-\lambda\kappa$ flips the ordering, while raising both sides to the power $\lambda\kappa$ does not and finally one finds
\begin{equation}\label{eq: RegionCondForLambdaLargerThanThreeOverTwo}
	\frac{N}{N_0}
	\geq
	 \left(\frac{2}{3\kappa}\right)^{\!\!\lambda\kappa} \exp{\frac{3}{2}\kappa}
	\quad
	0 < \lambda \leq \frac{3}{2}, \quad \kappa > 0\ep
\end{equation}

Additionally we should satisfy the reality constraint in \eqref{eq: RealityCondForLambdaKappaPos}, namely:
\begin{equation}\label{eq: RealityCondForLambdaLargerThanThreeOverTwo}
	\frac{N}{N_0}
	\geq
	 \left(\frac{1}{\lambda\kappa}\right)^{\!\!\lambda\kappa} \exp{\lambda\kappa} \ep
\end{equation}

We notice that for $\lambda=\frac{3}{2}$, Eq.~\eqref{eq: RealityCondForLambdaLargerThanThreeOverTwo} reduces to Eq.~\eqref{eq: RegionCondForLambdaLargerThanThreeOverTwo}, while it is easy to verify that $\forall \kappa>0$:
\begin{equation}
	\left(\frac{\text{e}}{\lambda\kappa}\right)^{\!\!\lambda\kappa}
	\leq
	\left(\frac{2}{3\kappa}\right)^{\!\!\lambda\kappa} \exp{\frac{3}{2}\kappa} \ec
\end{equation}
hence we can safely assume that the reality constraint is automatically satisfied if we impose Eq.~\eqref{eq: RegionCondForLambdaLargerThanThreeOverTwo}.

\vspace*{1.5\baselineskip}
\noindent {$\boxed{\lambda > 3/2}$}
\vspace*{0.5\baselineskip}

This implies that the \rhs of \eqref{eq: TempRegionCondForLambdaPos} is larger than $-1$, which does not set any constraint on the \lhs If we choose the real branch $\mathcal{W}_{-1}$, the \lhs will be smaller than $-1$ and hence ordering \eqref{eq: ProductExpOrdering4} applies, and we do not need to simplify \eqref{eq: TempRegionCondForLambdaPos} any further as it is automatically satisfied. In this case the reality constraint in Eq.~\eqref{eq: RealityCondForLambdaLargerThanThreeOverTwo} is the only one that remains. 

On the other hand, if we choose the real branch $\mathcal{W}_0$, the \lhs will be larger than $-1$ and hence ordering \eqref{eq: ProductExpOrdering1} applies, and we can simplify \eqref{eq: TempRegionCondForLambdaPos} into
\begin{equation}
	-\frac{1}{\lambda\kappa} \sqrt[\leftroot{-3}\uproot{3}\lambda\kappa]{\frac{N_0}{N}}
	\leq
	-\frac{3}{2\lambda} \exp{-\frac{3}{2\lambda}}
	\quad
	\lambda > \frac{3}{2}, \quad \kappa >0 \ep
\end{equation}

Multiplying both sides with $-\lambda\kappa$ changes the ordering, while raising both sides to the power $\lambda\kappa$ does not and finally one finds
\begin{equation}
	\frac{N}{N_0}
	\leq
	\left(\frac{2}{3\kappa}\right)^{\!\!\lambda\kappa} \exp{\frac{3}{2}\kappa}
	\quad
	\lambda > \frac{3}{2}, \quad \kappa >0 \ep
\end{equation}

However, we also have to satisfy the reality condition, which combined with the previous constraint, provides bounds from below and from above to $N$, namely
\begin{equation}
	\left(\frac{\text{e}}{\lambda\kappa}\right)^{\!\!\lambda\kappa}
	\leq
	\frac{N}{N_0}
	\leq
	\left(\frac{2}{3\kappa}\right)^{\!\!\lambda\kappa} \exp{\frac{3}{2}\kappa}
	\quad
	\lambda > \frac{3}{2}, \quad \kappa >0 \ep
\end{equation}

It is clear that we do not want an upper bound for $N/N_0$ as this would limit the validity of the Nekhoroshev stability-time estimate. Therefore we conclude that only the branch $\mathcal{W}_{-1}$ should be used in our application and the summary of all possible parameters' values is listed in Table~\ref{tab: ParameterValues}.


\begin{thebibliography}{99}
%
\bibitem{tev1} N.~M.~Gelfand, Calculations of the Dynamic Aperture at the Tevatron, in Proceedings of SSC Workshop on Accelerator Physics Issues for a Superconducting Super Collider, edited by M.~Tigner, 
UM-HE-84-1, 124, 1984.
%
\bibitem{tev2} V.~Visnjic, Dynamic aperture of low beta lattices at Tevatron collider, in Proceedings of 1991 Particle Accelerator Conference, edited by J.~Chew and  L.~Lizama (IEEE Computer Society Press, Piscataway -NY), 1701, 1991.
%
\bibitem{tev3} V.~Visnjic, Dynamic aperture of the future Tevatron Collider,  in Proceedings of Workshop On Nonlinear Problems In Future Particle Accelerators, edited by W.~Scandale and G.~Turchetti (World Scientific, Teaneck, NJ), 1991.
%
\bibitem{herap1} R.~Brinkmann, F.~Willeke, Persistent Current Field Errors and Dynamic Aperture of the Hera Proton Ring, DESY-HERA-88-08, 1988.
%
\bibitem{herap2} F.~Zimmermann, F.~Willeke, Long term stability and dynamic aperture of the HERA proton ring, 
DESY-HERA-91-08, 1991.
%
\bibitem{herap3} F.~Zimmermann, Dynamic aperture and transverse proton diffusion in HERA, SLAC-PUB-6458, 1994.
%
\bibitem{rhic} Y.~Luo, M.~Bai, J.~Beebe-Wang, W.~Fischer, A.~Jain, C.~Montag, T.~Roser, S.~Tepikian, D.~Trobjevic, Dynamic aperture evaluation at the current working point for RHIC polarized proton operation, in Proceedings of 2007 Particle Accelerator Conference, edited by C.~Petit-Jean-Genaz (IEEE Computer Society Press, Piscataway - NY, 2007), 4363, 2007.
%
\bibitem{LHCDR} O.~Br\"{u}ning, P.~Collier, Ph.~Lebrun, S.~Myers, R.~Ostojic, J.~Poole, P.~Proudlock (eds.), LHC Design Report, Vol.~1, CERN-2004-003-V-1, 2004.
%
\bibitem{fut1} S.~Tygier, R.B.~Appleby, J.M.~Garland, H.~Owen, J.~Pasternak, J-B.~Lagrange, Dynamic Aperture Studies of the nuSTORM FFAG RING, in Proceedings of 5th International Particle Accelerator Conference, ed. by C.~Petit-Jean Genaz, G.~Arduini, P.~Michel, V.R.W.~Schaa, 1574 (2014).
%
\bibitem{fut2} Y.~Jing, V.N.~Litvinenko, D.~Trbojevic, Optimization of Dynamic Aperture for Hadron Lattices In eRHIC, in Proceedings of 6th International Particle Accelerator Conference, ed. by S.~Henderson, T.~Satogata, V.R.W.~Schaa, 757 (2015).

\bibitem{fut3} B.~Dalena, D.~Boutin, A.~Chanc\'e, J.~Payet, B.~Holzer, R.~Martin, D.~Schulte, First Evaluation of Dynamic Aperture at Injection for FCC-hh, in Proceedings of 7th International Particle Accelerator Conference, ed. by K.S.~Kim, C.~Petit-Jean-Genaz, I.S.~Ko, K.R.~Kim, V.R.W.~Schaa, 1466 (2016).
%
\bibitem{fut4} B.~Dalena, D.~Boutin, A.~Chanc\'e, B.~Holzer, D.~Schulte, Advance on Dynamic Aperture at injection For FCC-hh, in Proceedings of 8th International Particle Accelerator Conference, ed. by V.R.W~Schaa, G.~Arduini, M.~Lindroos, J.~Pranke, 2027 (2017).
%
\bibitem{fut5} E.~Cruz-Alaniz,A.~Seryi, E.H.~Maclean, R.~Martin, R.~Tom\'as, Non Linear Field Correction Effects on the Dynamic Aperture of the FCC-hh, in Proceedings of 8th International Particle Accelerator Conference, ed. by V.R.W~Schaa, G.~Arduini, M.~Lindroos, J.~Pranke, 2143 (2017).
%
\bibitem{TDR} High-Luminosity Large Hadron Collider (HL-LHC). Technical Design Report V.0.1, edited by G.~Apollinari, I.~Bejar Alonso, O.~Br\"uning, P.~Fessia, M.~Lamont, L.~Rossi, L. Tavian, CERN Yellow Reports: Monographs, Vol.4/2017, CERN-2017-007-M (CERN, Geneva, 2017). https://doi.org/10.23731/CYRM-2017-004.
%
\bibitem{fut6} B.~Dalena, D.~Boutin, A.~Chanc\'e, B.~Holzer, S.~Izquierdo Bermudez, D.~Schoerling, D.~Schulte, Dipole Field Quality and Dynamic Aperture for FCC-hh, in Proceedings of 9th International Particle Accelerator Conference, ed. by S.~Koscielniak, T.~Satogata, V.R.W.~Schaa, J.~Thomson, 137 (2018).
%
\bibitem{fut7} E.~Cruz-Alaniz, J.L.~Abelleira, L.~van Riesen-Haupt, A.~Seryi, R.~Martin, R.~Tom\'as, Methods to Increase the Dynamic Aperture of the FCC-hh LATTICE, in Proceedings of 9th International Particle Accelerator Conference, ed. by S.~Koscielniak, T.~Satogata, V.R.W.~Schaa, J.~Thomson, 3593 (2018).
%
\bibitem{dynap1} M.~Giovannozzi, W.~Scandale, E.~Todesco, Prediction of long-term stability in large hadron colliders, Part. Accel. {\bf 56}, 195 (1996).
%
\bibitem{invlog} M.~Giovannozzi, W.~Scandale, E.~Todesco, Dynamic aperture extrapolation in presence of tune modulation, Phys. Rev. E {\bf 57}, 3432 (1998).
%
\bibitem{da_and_losses} M.~Giovannozzi, A proposed scaling law for intensity evolution in hadron storage rings based on dynamic aperture variation with time, Phys. Rev. ST Accel. Beams {\bf 15} 024001, (2012).
%
\bibitem{KAM1} A.N.~Kolmogorov, On the Conservation of Conditionally Periodic Motions under Small Perturbation of the Hamiltonian, Dokl. Akad. Nauk SSR 98 (1954).
%
\bibitem{KAM2} J.~Moser, On invariant curves of area-preserving mappings of an annulus, Nachr. Akad. Wiss. Göttingen Math.-Phys. Kl. II 1 (1962).
%
\bibitem{KAM3} V.I.~Arnold, Proof of a theorem of A.N.~Kolmogorov on the preservation of conditionally periodic motions under a small perturbation of the Hamiltonian, Russ. Math. Surv. {\bf 18} 9 (1963).
%
\bibitem{KAM4} C.~L.~Siegel and J.~Moser, {\sl Lectures in celestial
  mechanics}, Berlin Springer Verlag, 1971.
%
\bibitem{Nekhoroshev:1971aa} N.~Nekhoroshev, Behavior of Hamiltonian systems close to integrable, Functional Analysis and Its Applications {\bf 5} 338 (1971).
%
\bibitem{Nekhoroshev:1977aa} N.~Nekhoroshev, An exponential estimate of the time of stability of nearly-integrable Hamiltonian systems, Russ. Math. Surv. {\bf 32}, 1 (1977).
%
\bibitem{Bazzani:1990aa} A.~Bazzani, S.~Marmi, G.~Turchetti, Nekhoroshev estimate for isochronous non resonant symplectic maps, Cel. Mech. {\bf 47}, 333 (1990).
%
\bibitem{Turchetti:1990aa} G.~Turchetti, Nekhoroshev stability estimates for symplectic maps and physical applications, in {\em Proceedings of Number Theory and Physics}, edited by J.~M.~Luck, P.~Moussa, M.~Waldschmidt, (Berlin Springer Verlag) Springer Proceedings in Physics, V. 47, 223 (1990).
%
\bibitem{LumiI} M.~Giovannozzi, F. Van der Veken, Description of the luminosity evolution for the CERN LHC including dynamic aperture effects, Part I: The model, Nucl. Instrum. \& Methods A {\bf 905}, 171.
%
\bibitem{LumiII} M.~Giovannozzi, F. Van der Veken, Description of the luminosity evolution for the CERN LHC including dynamic aperture effects. Part II: application to Run 1 data, Nucl. Instrum. \& Methods A {\bf 908}, 1.
%
\bibitem{Knuth} R.M.~Corless, G.H.~Gonnet, D.E.G.~Hare, D.J.~Jeffrey and D.E.~Knuth, On the Lambert W Function, Adv. Comput. Math. {\bf 5}, 329 (1996).
%
\bibitem{FZ} M.~Benedikt, D.~Schulte, and F.~Zimmermann, Optimizing integrated luminosity of future hadron colliders, Phys. Rev. ST Accel. Beams {\bf 18}, 101002 (2015). 
%
\bibitem{Knuth1} R.L.~Graham, D.E.~Knuth, O.~Patashnik, {\em Concrete Mathematics}, Addison Wesley (1994).
%
\bibitem{Knuth2} D.J.~Jerey, R.M.~Corless, D.E.G Hare, D.E.~Knuth, Sur l'inversion de $y^\alpha e^y$ au moyen de nombres de Stirling associ\'es, C. R. Acad. Sc. Paris, S\'erie I, {\bf 320}, 1449 (1995).  
%
\bibitem{yell} A.~Bazzani, E.~Todesco, G.~Turchetti and G.~Servizi, A normal form approach to the theory of nonlinear betatronic motion, CERN Yellow report 94--02 (1994).
%
\bibitem{dacomp} E.~Todesco and M.~Giovannozzi, Dynamic aperture estimates and phase-space distortions in nonlinear betatron motion, Phys. Rev. E {\bf 53}, 4067 (1996). 
%
\bibitem{DABeam2} E.~H.~Maclean, R.~Tom\'as, F.~Schmidt, and T.~H.~B.~Persson, Measurement of nonlinear observables in the Large Hadron Collider using kicked beams, Phys. Rev. ST Accel. Beams {\bf 17}, 081002 (2014). 
%
\bibitem{DABeam1_1} M.~Albert, G.~Crockford, S.~Fartoukh, M.~Giovannozzi, E.~Maclean, A.~MacPherson, R.~Miyamoto, L.~Ponce, S.~Redaelli, H.~Renshall, F.~Roncarolo, R.~Steinhagen, E.~Todesco, R.~Tom\'as, W.~Venturini Delsolaro, First Experimental observations from the LHC Dynamic Aperture Experiment, in {\em Proceedings of the Third International Particle Accelerator Conference}, edited by C.~Eyberger and F.~Zimmermann, 1362 (2012).
%
\bibitem{DABeam1_2} M.~Giovannozzi, S. Cettour Cave, R. De Maria, M. Ludwig, A. Macpherson, S. Redaelli, F. Roncarolo, M. Solfaroli Camillocci, W. Venturini Delsolaro, Experimental Observations from the LHC Dynamic Aperture Machine Development Study in 2012, in {\em Proceedings of the Fourth International Particle Accelerator Conference}, edited by Z.~Dai, C.~Petit-Jean-Genaz, V.~R.~W. Schaa, C.~Zhang, 2606 (2013).
%
%
\bibitem{MD_note} E.H.~Maclean, F.~Carlier, M.~Giovannozzi, R.~Tom\'as, Report from LHC MD 2171: Dynamic aperture at 6.5~TeV, CERN-ACC-Note-2018-0054.
%
\bibitem{DAasbuilt} S.~Fartoukh, M.~Giovannozzi, ``Dynamic aperture computation for the as-built CERN Large Hadron Collider and impact of main dipoles sorting'', Nucl. Instrum. \& Methods A {\bf 671} 10 (2012).
%
\bibitem{sixtrack} http://sixtrack.web.cern.ch/SixTrack/
%
\end{thebibliography}
\end{document}